\documentclass{aa}

\usepackage{graphicx}
\usepackage[normalem]{ulem}
\usepackage{txfonts}
\usepackage[modulo,switch]{lineno}
\usepackage{hyperref}  
\hypersetup{
    colorlinks=true,
    linkcolor=blue,
    filecolor=magenta,
    citecolor=blue,
    urlcolor=blue,
}
\usepackage{natbib}
\bibpunct{(}{)}{;}{a}{}{,}
\usepackage{multirow}
\usepackage{nicefrac}
\usepackage[usenames,dvipsnames]{color}

\begin{document}

\title{Determining the dynamics and magnetic fields in He\,\textsc{i} 10830 \AA\ 
during a solar filament eruption}
\author{C. Kuckein\inst{1}, S. J. Gonz\'alez Manrique\inst{2,3,4}, L. Kleint\inst{5}, and 
A. Asensio Ramos\inst{3,4}}

\authorrunning{Kuckein et al.}

\institute{%
    $^1$ Leibniz-Institut f{\"u}r Astrophysik Potsdam (AIP),
         An der Sternwarte 16, 
         14482 Potsdam, Germany\\
    $^2$ Astronomical Institute, Slovak Academy of Sciences (AISAS), 
         05960 Tatransk\'{a} Lomnica, Slovak Republic\\
    $^3$ Instituto de Astrof\'{i}sica de Canarias (IAC), 
         V\'{i}a L\'{a}ctea s/n, 38205 La Laguna, Tenerife, Spain\\     
    $^4$ Departamento de Astrof\'{\i}sica, Universidad de La Laguna
         38205, La Laguna, Tenerife, Spain \\ 
    $^5$ Leibniz-Institut f{\"u}r Sonnenphysik (KIS), 
         Sch\"oneckstrasse 6, 79104 Freiburg im Breisgau, Germany\\
    \email{ckuckein@aip.de}}

\date{Received \today; accepted later}

\date{\today}

\abstract{% context (optional)
}
{% aims
We investigate the dynamics and magnetic properties of the plasma, such as line-of-sight velocity (LOS), 
optical depth, vertical and horizontal magnetic fields, belonging to an erupted solar filament.
}
{% methods
The filament eruption was observed with the GREGOR Infrared Spectrograph
(GRIS) at the 1.5-meter GREGOR telescope on 2016 July 3. 
Three consecutive full-Stokes slit-spectropolarimetric scans in the
\ion{He}{i} 10830\,\AA\ spectral range were acquired. The Stokes $I$ profiles were classified using
the machine learning $k$-means algorithm and then inverted with different 
initial conditions using the HAZEL code. }
{% results
The erupting-filament material presents the following physical conditions: 
(1) ubiquitous upward motions with peak LOS velocities of $\sim$73\,km\,s$^{-1}$; 
(2) predominant large horizontal components of the magnetic field, on average, 
in the range of 173--254\,G, whereas
the vertical components of the fields are much lower, on average between 39--58\,G; 
(3) optical depths in the range of 0.7--1.1.
The average azimuth orientation of the field lines between two consecutive raster scans (<2.5 minutes) 
remained constant. 
}
{% conclusion (optional)
The analyzed filament eruption belonged to the fast rising phase, with total velocities of
about 124\,km\,s$^{-1}$. 
The orientation of the magnetic field lines does not change from one 
raster scan to the other, indicating that the untwisting phase
has not started yet. The untwisting seems to start about 15\,min after the beginning
of the filament eruption. 
}

  \keywords{Sun: filaments --
             Sun: chromosphere --
             Sun: magnetic fields --
             Methods: data analysis --
             Techniques: high angular resolution --
             Techniques: polarimetric}

\maketitle

%--------------------------------------------------------------------------
\section{Introduction}
%--------------------------------------------------------------------------
Filaments belong to the largest structures seen in the chromosphere
and corona of the Sun. They are composed of fine threads of plasma which are confined
by magnetic field lines. Filaments, also called prominences when observed off-disk above 
the limb, are easily identified using spectral
lines that map the chromosphere, for example,  H$\alpha$ or the \ion{He}{i} 10830\,\AA\ triplet. 
Quiescent filaments are large and long-lived phenomena in the solar atmosphere. 
They are located outside of active 
regions and have typically lifetimes from days to weeks. 
On the contrary, active region (AR) filaments are always found inside active regions, 
are smaller in size, and more dynamic than their quiescent counterpart. 
Their lifetimes are usually between hours and days. 
For a review on the main properties of filaments, the reader is 
referred to, for example, \citet{tandberg95}, \citet{mackay10}, \citet{parenti14}, and \citet{vial15}.

Filaments are found
on top of photospheric polarity inversion lines \citep{babcock55}
and it is commonly accepted that they are sustained 
against gravity by the magnetic field lines. The topology of these 
field lines is often reported as a flux rope,
where field lines twist around the plasma at the main axis or spine of the filament. There are 
numerous studies supporting this scenario by means of models 
\citep[see, e.g.,][]{vanballe89, amari99}, extrapolations \citep[see, e.g.,][]{guo10,canou10,jing10,yelles12},
and observations \citep[see, e.g.,][]{kuckein12a,xu12,wang20}.
Magnetic field strengths of 20--40\,G are typically found in quiescent filaments
\citep[see, e.g.,][]{trujillo02, merenda06}, whereas AR filaments can host fields
up to 600--800\,G \citep{kuckein09, guo10, xu12}. However, we point out
that AR filaments might also harbor weaker fields \citep{diazbaso16}
because they are partially transparent to the radiation coming
from the active region below. This severely complicates the inference 
of magnetic fields in AR filaments \citep{diaz_baso19a,diaz_baso19b}.
Disruptions of the magnetic field can produce the loss of 
equilibrium of the filament's structure producing an eruption. For instance, 
reconnection of the magnetic field lines lead to dramatic changes in the magnetic
field configuration, involving the release of energy and opening of field lines
\citep[see, e.g.,][]{antiochos99, moore01}.
Conversely, magnetic reconnection can also be a consequence of
ideal magnetohydrodynamic (MHD) instabilities \citep{low96}. MHD simulations
have shown that instabilities, such as the kink instability \citep{torok05}, 
can also trigger large-scale eruptions. Also, a torus instability can lead to
a flux rope eruption \citep{kliem06}.

In a statistical study based on 106 major filament-eruption events observed at the 
Big Bear Solar Observatory (BBSO), \citet{jing04} concluded 
that a bit more than half of the events were associated with coronal mass ejections (CMEs). 
Furthermore, AR filament eruptions are more likely associated with flares (95\,\%) than quiescent 
filament eruptions (27\,\%). In their study, 64\,\% of the disk events were related to 
new flux emergence. The study revealed that different mechanisms may be responsible for the 
triggering of filament eruptions. Similar eruptive events, such as coronal mass ejections (CMEs),
have been associated with both, an increase and/or decrease of photospheric 
magnetic flux \citep{zhang08}. 
Flux cancellation plays an important role for low-altitude eruptions, but additionally 
requires the removal of the overlying containing field \citep{yardley18}.

In the present study we will infer the plasma properties of an ejected filament.
From individual case studies, using the \ion{He}{i} 10830\,\AA\ triplet, it has been reported
that ejected plasma can reach line-of-sight (LOS) velocities between 200--300\,km\,s$^{-1}$ 
\citep{penn00}. An additional difficulty lies in the determination of physical quantities under
such strong spectral-line shifts, which typically come along with multiple displaced Doppler 
components \citep[see, e.g.,][]{muglach97,penn95,gonzalez16}.
\citet{sasso11, sasso14} needed up to five different atmospheric components to reproduce the observed 
\ion{He}{i} line profiles with an inversion code and 
to extract the LOS velocities and magnetic field properties. 
In their study, they tracked an activated filament in a flaring environment, 
showing upflows of up to 60\,km\,s$^{-1}$. Moreover, their blueshifted spectral-line components
were mainly associated with transverse magnetic fields in the body of the filament.
Multiple Gaussian fits were also necessary to fit the H$\alpha$-line profiles in a filament 
eruption observed with the Swedish Solar Telescope (SST, La Palma, Spain) by \citet{doyle19}. Their 
strongly blueshifted H$\alpha$ profiles reached at least 60\,km\,s$^{-1}$, although the real
velocities were likely higher, but outside of the spectral range of the instrument.  
Recently, \citet{wang20} followed a quiescent filament eruption with the Dunn Solar 
Telescope (DST, New Mexico, USA). The authors concentrated on the magnetic properties of the filament and 
found homogeneous linear polarization signals in the \ion{He}{i} 10830\,\AA\ triplet, 
which they interpret as a magnetic flux rope. 

Here we report on the eruption of a quiescent filament where the expelled plasma crossed the 
slit spectrograph of the 1.5-meter \mbox{GREGOR} telescope \citep{schmidt12} 
located on the island of Tenerife, Spain. We perform an analysis of the four 
Stokes profiles arising from the chromospheric \ion{He}{i} 10830\,\AA\ triplet 
detected during the eruption process. The goal is to characterize the 
ejected plasma of the filament.

%--------------------------------------------------------------------------
\section{Observations}
%--------------------------------------------------------------------------

\begin{figure}[!t]
 \centering
  \includegraphics[width=\hsize]{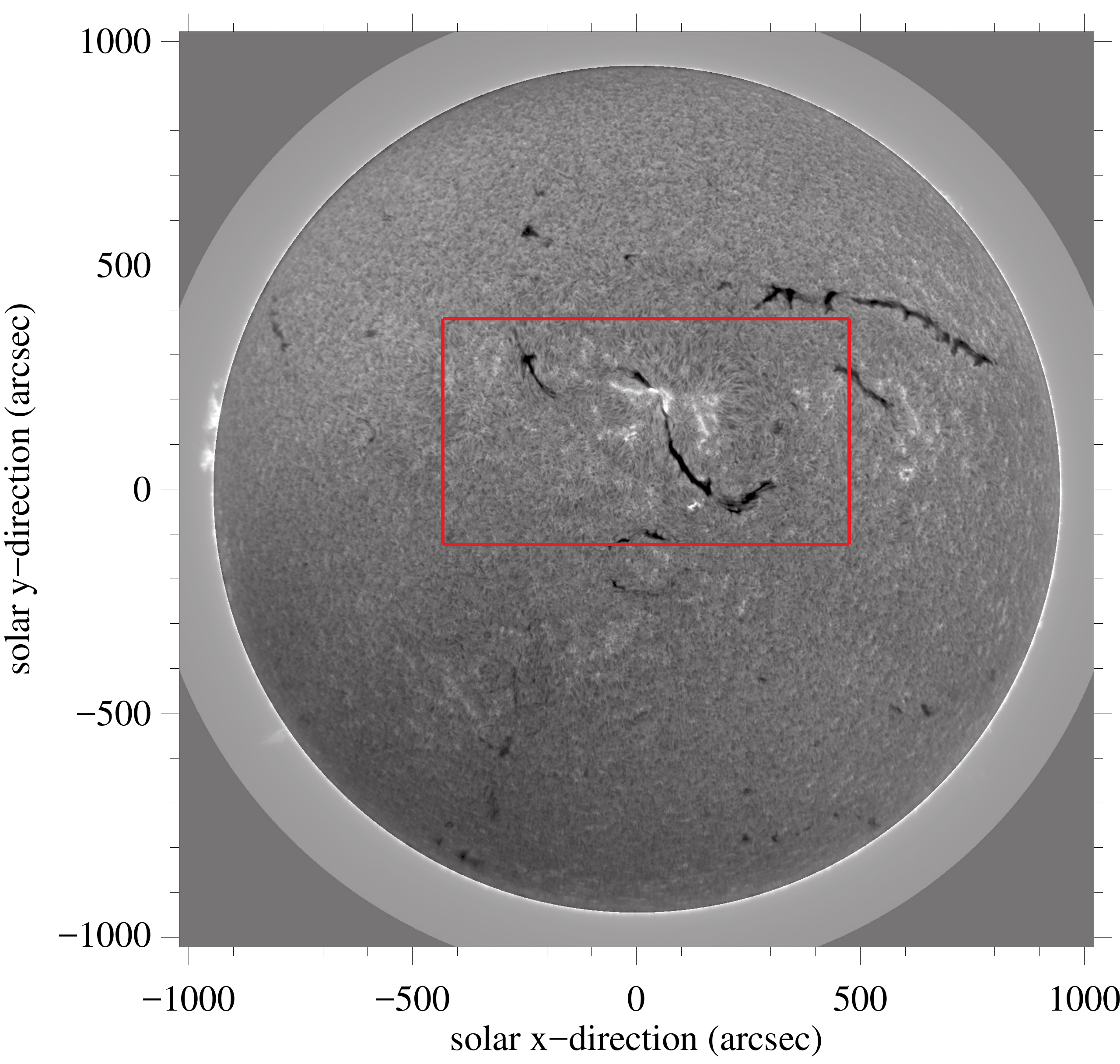} 
\caption{ChroTel H$\alpha$ filtergram on 2016 July 3 at 09:39 UT before the eruption. 
Solar north is up and solar west is 
right. The red rectangle outlines
the region-of-interest shown in Fig. \ref{Fig:HMIfilament}. An animation of the ChroTel 
images during the filament eruption is available as an online movie. } 
 \label{Fig:Chrotel}
 \end{figure}

\begin{figure}[!t]
 \centering
 \includegraphics[width=\hsize]{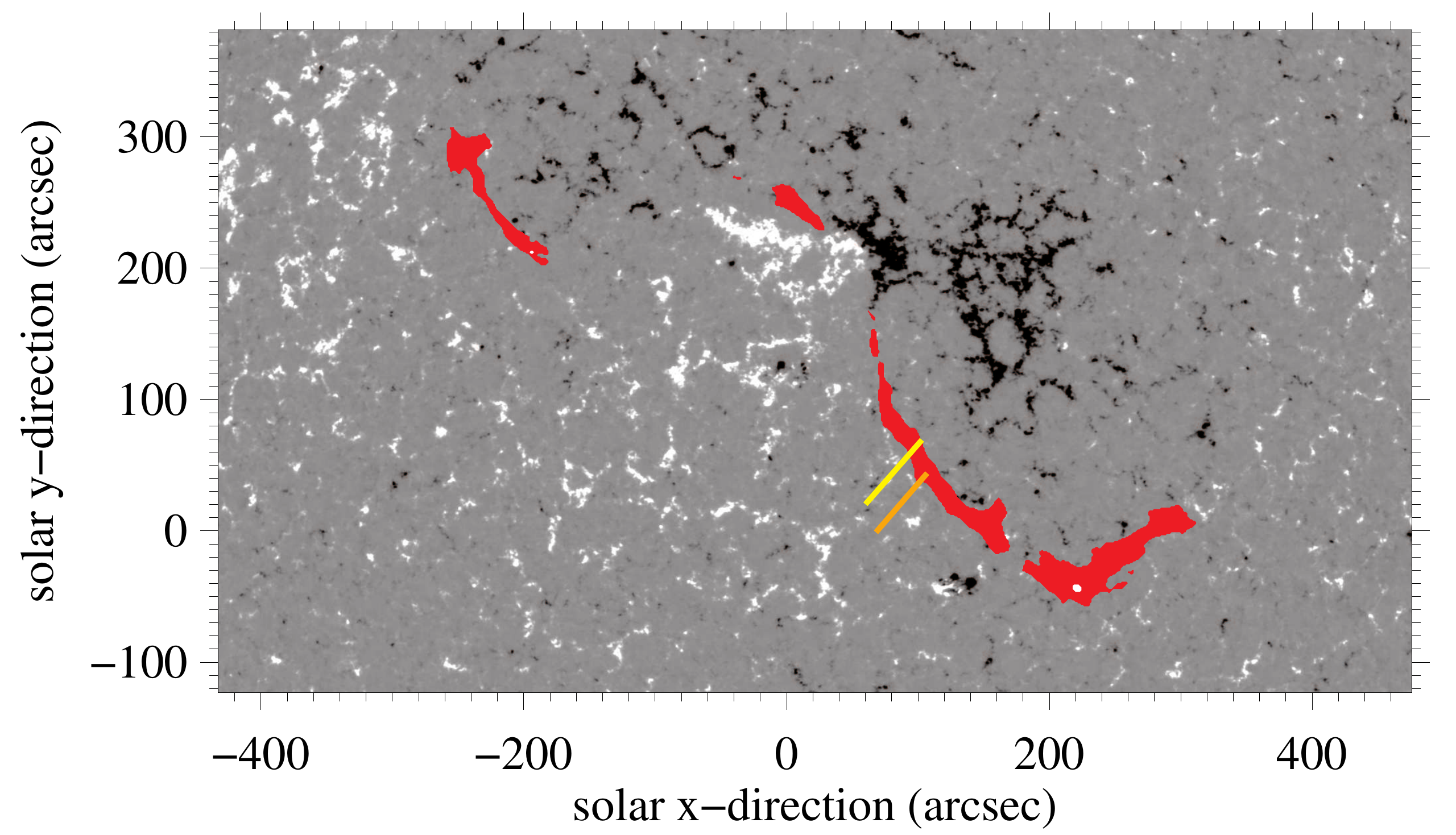} 
\caption{Deep HMI magnetogram of 160 summed up de-rotated individual magnetograms 
         between 09:00 and 11:00 UT on
         2016 July 3. The magnetogram is clipped between $\pm$\,100\,G to enhance weak magnetic fields. 
         The filled red contour corresponds to the filament in H$\alpha$ at rest extracted from a ChroTel 
         filtergram before the eruption at 09:39\,UT. The yellow and orange lines
         represent the two slit positions of the spectrograph at about 10:06\,UT and 10:09\,UT, 
         corresponding to maps A--C.} 
  \label{Fig:HMIfilament}
\end{figure}

\begin{figure*}[!t]
 \centering
 \includegraphics[width=\textwidth]{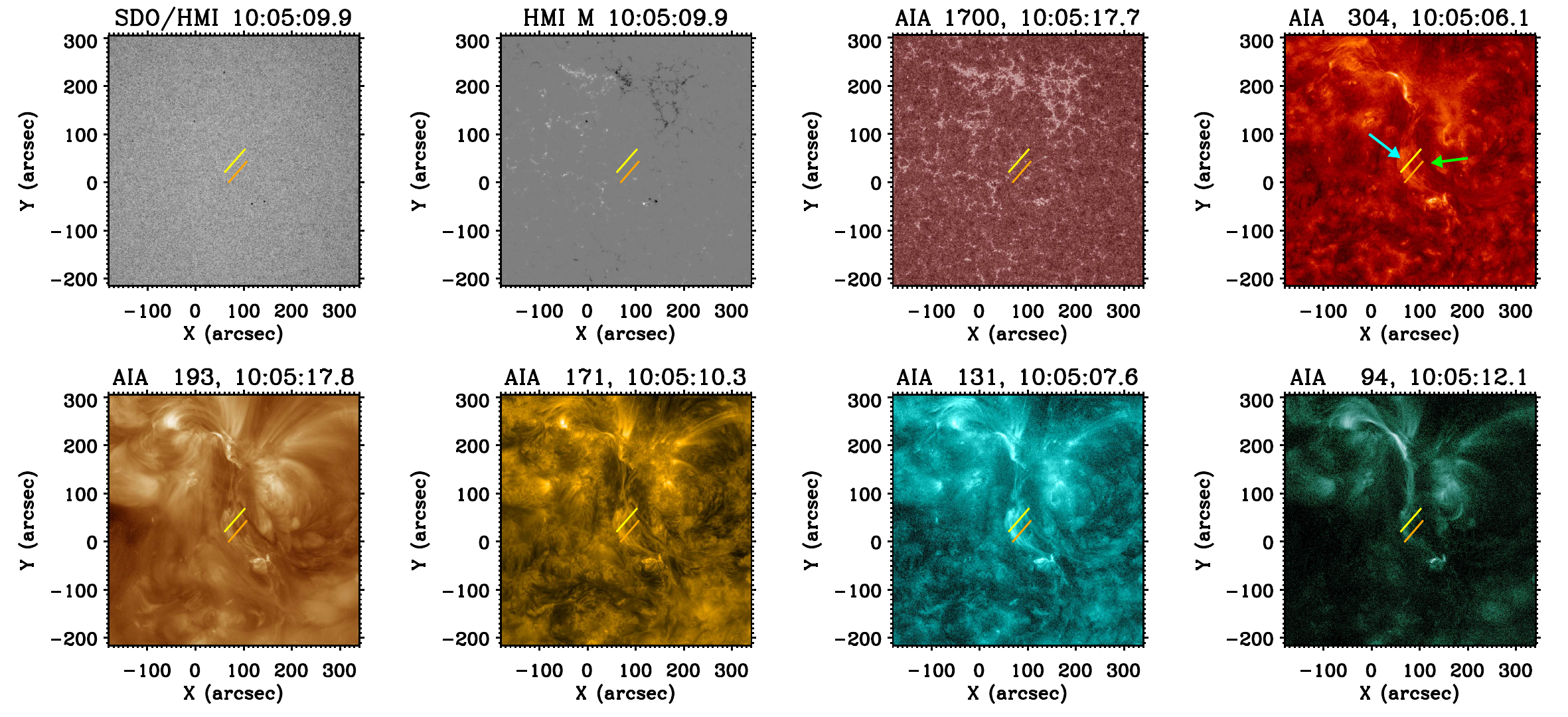} 
\caption{SDO overview images with an overlap of the slit for maps A and B (yellow line, at about 10:06\,UT), 
and map C (orange line at about 10:09\,UT) from the spectrograph at GREGOR. 
Clockwise starting top left: HMI continuum and magnetogram followed by several AIA channels. 
The cyan and green arrows in the upper right 304\,\AA\ filtergram point to the erupted and stable 
filament, respectively.
This figure is available as an online movie between 08:30\,UT and 13:00\,UT. } 
  \label{Fig:AIAoverview}
\end{figure*}

\subsection{Ground-based observations}
The filament eruption was observed on 2016 July 3 with the ground-based \mbox{GREGOR} 
telescope and the full-disk imager Chromospheric Telescope 
\citep[ChroTel,][]{kentischer08,bethge11} both located at the Observatorio del Teide, Tenerife, Spain. 
To our knowledge, this is the first filament eruption observed by GREGOR.
Figure \ref{Fig:Chrotel} shows an H$\alpha$ full-disk image of the Sun before the eruption started.
Full-Stokes slit-spectropolarimetry with the GREGOR
Infrared Spectrograph \citep[GRIS,][]{collados12} was acquired close to disk center ($\mu$=0.99) 
during the eruption of the filament. 
Three short raster scans of 20 steps each during the maximum of the event 
assured a proper tracking of the eruption (see Table \ref{tab:gris}). 
The first two raster scans, A and B, were taken between 
10:02\,UT and 10:07\,UT, whereas the third one C was recorded between 10:08\,UT and 10:10\,UT, with
a slightly shifted field-of-view (FOV) toward South with respect to the first two scans. 
The step size of the slit was 0\farcs135 and each step consisted of ten accumulations with an exposure
time of 100\,ms each. The spatial sampling along the slit was 0\farcs136 and the orientation is shown 
as yellow (maps A and B) and orange (map C) lines in 
Figs. \ref{Fig:HMIfilament} and \ref{Fig:AIAoverview}. 
The GREGOR polarimetric unit \citep{hofmann12} was used to carry out the polarimetric calibration
of the data. Dark and flat-field corrections were performed following the standard procedures
described by \citet{collados99, collados03}. The observations significantly 
benefited from the adaptive optics system \citep{Berkefeld10} which was locked on granulation.
According to the open data policy of GREGOR and SOLARNET, the observations can be freely downloaded
from the GRIS data archive\footnote{http://sdc.leibniz-kis.de}. 

\begin{table}[!hb]
\begin{center}
\caption{Spectropolarimetric raster scans of GRIS at GREGOR on 2016 July 3.
The effective exposure time in the last column is the number of accumulations times the
individual exposure time. }\label{tab:gris}
\begin{tabular}{clll}
\hline
\hline
Map             & Time (UT)              & Steps & Eff. t$_\mathrm{exp}$ (ms) \rule[-4pt]{0pt}{15pt}  \\
\hline
\hline
A               &   10:02:47 -- 10:04:36 &  20   & 1000 \rule[-4pt]{0pt}{14pt} \\
B               &   10:05:11 -- 10:07:00 &  20   & 1000 \\
C               &   10:08:12 -- 10:10:01 &  20   & 1000 \\
\hline
\end{tabular}
\end{center}
\end{table}

The spectral range spanned between 10824.3\,\AA\ and 10842.5\,\AA, with a spectral sampling of
18\,m\AA\,px$^{-1}$. This wavelength range includes, among other lines, the
photospheric \ion{Si}{i} 10827\,\AA\
line and the \ion{He}{i} 10830\,\AA\ triplet, the latter being formed in the upper chromosphere 
\citep{avrett94}. The triplet comprises the so-called ``blue'' component at 
10829.09\,\AA\ ($J_\mathrm{L}$$=$1 $\rightarrow$ $J_\mathrm{U}$$=$0) and
a blended ``red'' component at $\sim$10830.30\,\AA\ ($J_\mathrm{L}$$=$1 $\rightarrow$ $J_\mathrm{U}$$=$1, 2).

In addition to the high-resolution GREGOR images, the auxiliary ChroTel was running in a non-standard 
observing mode, acquiring full-disk H$\alpha$
images with a higher cadence of one minute. ChroTel is a small full-disk imager located on the flat roof of the 
Vacuum Tower Telescope \citep[VTT,][]{vonderluehe98} next to the GREGOR telescope. The images
were corrected for dark current, flat-field, and limb darkening. 
An overview image of the scene with ChroTel before the eruption started is shown in 
Fig. \ref{Fig:Chrotel}. 
The full time series, including the filament eruption, is available as an online movie.

%---- Figure X -----------------------------------------------------------------
\begin{figure*}[!t]
\sidecaption
  \includegraphics[width=12cm]{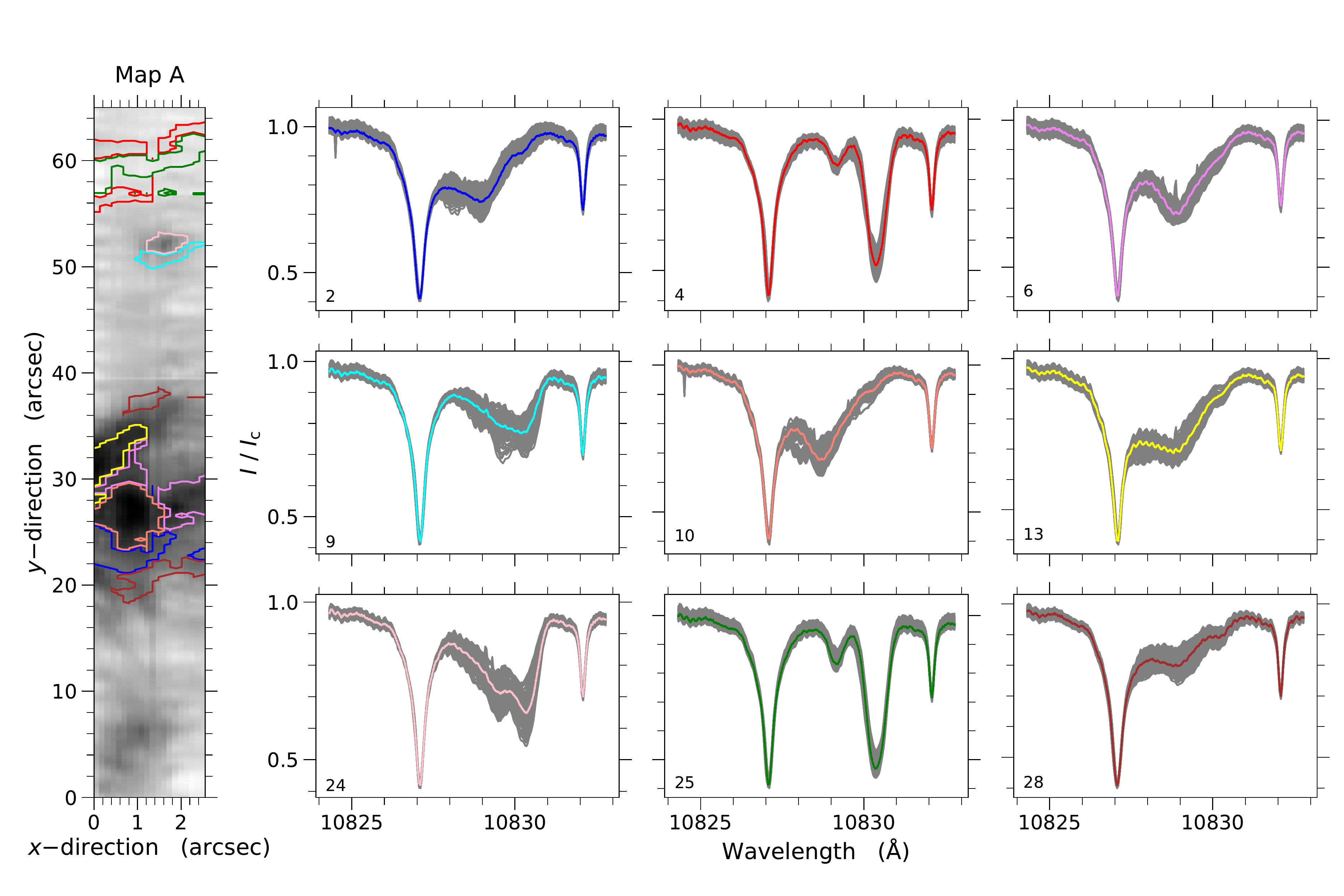} 
  \caption{Selected groups of similar Stokes $I$ profiles determined using $k$-means classification. 
  \emph{Left}: Slit-reconstructed \ion{He}{i} image centered 
  at 10828.5 \AA, in the blue wing of \ion{He}{i}, corresponding to a Doppler shift of 
  $-$50\,km\,s$^{-1}$ of the \ion{He}{i} red component.
  The vertical direction is 
  along the slit. The horizontal direction is the raster-scan direction. The dark pixels 
  represent blueshifted intensity profiles related to the erupted filament. 
  The profiles inside the color-coded contours are shown in the right-hand panels.
  \emph{Right}: the colored spectral profiles show the average intensity profile of each group, 
  whereas the gray profiles depict all individual profiles of each group. } 
  \label{Fig:cluster9}
\end{figure*}
%-------------------------------------------------------------------------------

\subsection{Space-borne observations}
Filtergrams of the Atmospheric Imaging Assembly \citep[AIA,][]{aia} and magnetograms 
of the Helioseismic and Magnetic Imager \citep[HMI,][]{hmi} on board of the Solar 
Dynamics Observatory \citep[SDO,][]{sdo} were used as large context images.
They were aligned and calibrated using aia\_prep and they show the scene in the 
different layers from the photosphere to the corona (Fig. \ref{Fig:AIAoverview}).

%--------------------------------------------------------------------------
\section{Data analysis}
%--------------------------------------------------------------------------

\subsection{Alignment of SDO and GREGOR data}

For maps A and B, we aligned the GREGOR slitjaw data to 
the calibrated SDO data. We used simultanous data from 10:06\,UT 
taken by the slitjaw camera at GREGOR in the 7770\,\AA\ continuum and matched it to SDO/HMI continuum 
data by shifting and rolling the GREGOR image. This was performed manually and we estimate the 
precision to be within 0\farcs3. Because the slitjaw data show the spectrograph slit, 
its location and orientation could be retrieved easily and is shown as a yellow line in Figs. 
\ref{Fig:HMIfilament} and \ref{Fig:AIAoverview}. For map C, unfortunately, no slitjaw 
data were available, but there are simultaneous observations in a broadband channel at 
4307\,\AA\ with the High-resolution Fast Imager \citep[HiFI;][]{stools} instrument at GREGOR. 
We therefore had to carry out two steps to determine the slit alignment: 
(1) We matched the HiFI data at 08:31\,UT to SDO/HMI and also matched slitjaw data at 
that time to SDO/HMI, because all three instruments recorded at that time and 
because the target were pores, which simplified the alignment. In this way, we could 
determine the pixel coordinates to which the GRIS slit position corresponds in HiFI images. 
We then drew an artificial slit on the HiFI image at 10:09 (the time of map C) and aligned 
this HiFI image to SDO/HMI by rotating, resampling, and shifting. This allowed us to 
determine the slit position and orientation in solar coordinates. Because this method 
includes more steps, we estimate its precision to about 1\arcsec. The slit of map C 
is shown as an orange line in Figs. \ref{Fig:HMIfilament} and \ref{Fig:AIAoverview}.

\subsection{k-means clustering for the Stokes profiles}
A large variety of Stokes profiles with different shapes were found in the data.
This is naturally explained by the extreme atmospheric conditions present during eruptions
and flares. The presence of such different spectral profiles creates difficulties
when setting up the model and initial value of the parameters 
of an inversion code whose goal is to determine the physical 
parameters of the observed atmosphere. For spectral-line inversions, it is helpful to provide an
initial-guess atmosphere, which is fairly close to the real conditions. 
This avoids local minima, speeds up the inversion process, and improves correct convergence of the code.

%-------------------------------------------------------------------------------
\begin{figure*}[!t]
 \centering
 \includegraphics[width=\textwidth]{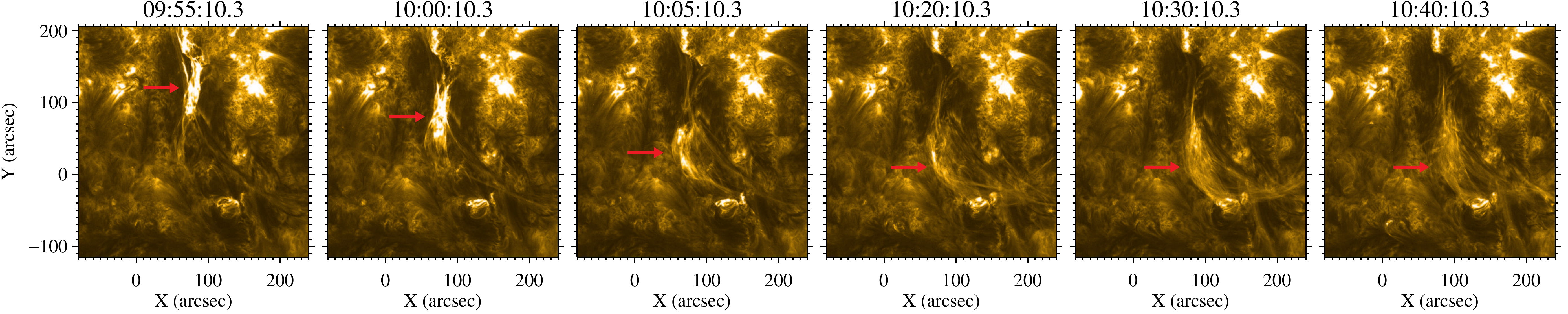}
\caption{AIA 171\,\AA\ time evolution of the filament eruption between 09:55--10:40\,UT. 
The red arrow marks the erupted material. 
The first three panels show a very dynamic rising phase of the filament, whereas 
the last three panels suggest a slower evolution, which is compatible with the
untwisting of field lines.} 
  \label{Fig:AIA171}
\end{figure*}
%-------------------------------------------------------------------------------

We used {\it k}-means clustering, an unsupervised machine learning algorithm, 
to classify the different types of Stokes $I$ profiles into similar groups. This technique 
is often used to group common spectral profiles in large data sets 
\citep[see, e.g.,][]{pietarila07, viticchie11,panos18, robustini19, sainz_dalda2019}. The {\it k}-means algorithm 
implementation used in this work is 
part of the \emph{scikit-learn} library for python.
We found that 30 groups cover well the diversity of the spectra.  
The average representative spectral profile for each
group is shown in Appendix \ref{App:A} (Fig. \ref{Fig:cluster30}). 
The gray profiles represent all individual profiles belonging to the same group. 
A selected group of relevant intensity profiles of map A is depicted in
\mbox{Fig. \ref{Fig:cluster9}}. Profiles corresponding to the central dark areas in the 
slit-reconstructed image, that are, groups 2, 6, 10, 13, and
28, have in common a strong blueshift of the \ion{He}{i} 10830\,\AA\ multiplet. 
These profiles are associated with the expelled plasma of the filament, the dark
areas in Fig. \ref{Fig:cluster9}. On the contrary, 
groups 4 and 25 show strong \ion{He}{i} absorption and are located where the filament
is found at rest. This is also the reason why the filament at rest is not seen in the upper 
part of the slit-reconstructed image, because only the blueshifted \ion{He}{i} is shown.

\subsection{HAZEL inversions}
One of the reasons for clustering the profiles into groups is to facilitate
their inversions with the inversion code HAZEL\footnote{We used HAZEL2, the current
version of the code can be found at \texttt{https://github.com/aasensio/hazel2}.}.
It is impossible to carry out an inversion of the whole FOV
with a single configuration for the input parameters. 
Profiles clearly showing only a single component in
the He \textsc{i} 10830 \AA\ multiplet should be inverted with a single 
chromospheric component. Likewise, those presenting more components should contain
the appropriate number of atmospheric components, so that the inversion code can
reliably determine the optical depths and velocities of the components.
Up to three atmospheric components were necessary to fit the extreme blueshifted
spectral profiles. 

The first task is to align the $Q$>$0$ reference directions in both the observations (defined
by the polarimeter) and HAZEL (defined by the $\gamma$ angle). To this end, 
we follow the description of the original paper \citep{hazel08} and
the examples shown in the manual\footnote{The
HAZEL manual, with a description of how to choose the reference direction in the
code can be found in \texttt{https://aasensio.github.io/hazel2/index.html}}.
The next step is the selection of atmospheric models that should match the observations.
These models contain the photosphere, treated in local thermodynamic
equilibrium (LTE), which produces
the \ion{Si}{i} 10827\,\AA\ line, and up to three chromospheres that produce the absorption in the
\ion{He}{i} 10830 \AA\ multiplets blended with the wing of the Si \textsc{i}
line and a very simple parameterized Voigt
function to fit the telluric line at $\sim$10832\,\AA. This last component is added to end up
with good fits in those cases in which the \ion{He}{i} triplet partially
blends with the telluric contamination due to a strong redshift. 
Concerning the photosphere, we
have found that 5 nodes in temperature, 1 node in microturbulent velocity and
2 nodes in bulk velocity produce sufficiently good fits.

Inversions are carried out in two cycles. The first cycle simply fits the
Stokes $I$ profile, with all the components in the \ion{He}{i} multiplet. 
The magnetic field is set to zero in this cycle. We found it necessary to 
put hard limits on the LOS velocity for each component to force the consistent 
isolation of all velocity components. In a second cycle, all Stokes
parameters are fitted by freeing the magnetic field in the photosphere
and all the chromospheres. As already put forward by \cite{diaz_baso19b},
the inversions of more than one single component in the \ion{He}{i}
10830 \AA\ multiplet leads to a panoply of ambiguous solutions. This is especially
relevant for cases in which all components have very similar velocities.
In those cases, the solutions we get are probably of limited interest.
However, this might not be the case when the components are clearly
segregated in velocity.

Contrary to the fitting of Stokes $I$, fitting the Stokes $Q$, $U$, and $V$ profiles 
turned out to be a complex task. After analyzing a few representative pixels, we
found that the weighting scheme $[5,5,1]$ for Stokes $Q$, $U$, and $V$ works
decently well for the cases with a single component, while the scheme
$[5,1/2,1]$ worked best for the cases with more than one component. The
reduced weight for Stokes $U$ is a consequence of the apparent incompatibility
of the synthetic Stokes $Q$ and $U$ signals with the observed ones. This is
an issue that requires a deep investigation and that we defer for the future. Reducing
the weight in Stokes $U$ produced a relatively good fit both in Stokes $Q$ and
$U$ on average.

%--------------------------------------------------------------------------
\section{Results}
%--------------------------------------------------------------------------

\subsection{Temporal evolution as seen by the full-disk instruments}

\subsubsection{Before the eruption}
We scrutinized H$\alpha$ full-disk images from the Kanzelh\"ohe Observatory (Austria)
data archive \citep{potzi13}, which
showed the filament for the first time on 2016 June 28. 
At the beginning, the filament channel is not completely filled, as several fragments of the 
spine do not show H$\alpha$ absorption. Over the next days, the fragments expand and 
merge to form a larger filament. 
It is worth to mention that the full-disk H$\alpha$ movies (not shown) exhibit that the
filament is very dynamic, that is, different pieces of the filament's spine merge, 
change their shape, split, etc, over the next days. 

As expected, the filament lays on top of the photospheric polarity inversion line (PIL) 
as depicted in Fig. \ref{Fig:HMIfilament}. Since the filament is very 
long, it is located on top of broader and narrower areas of the PIL. 
Of particular interest is the very narrow PIL 
located at ($x,y$$=60$\arcsec,220\arcsec). The HMI magnetogram (Fig. \ref{Fig:HMIfilament}) 
shows an ``abutted'' opposite polarity plage, which at some locations almost touch. 
Here the filament is more compact and confined.
The different AIA channels (\mbox{Fig. \ref{Fig:AIAoverview}}, see also the online movie) 
witness intensity enhancements,
arising from the compact PIL, starting at about 9:27\,UT. 
The enhancements are seen from the chromosphere 
(AIA 304\,\AA), via the transition region, up to the corona 
(AIA 171\,\AA, 193\,\AA, 94\,\AA, and 131\,\AA). 
Especially the channel at 131\,\AA\ shows five minutes later an increment 
of intensity south of the PIL, 
nearby the eastern part of the filament. The brightening occurs at the location where later 
the material of the erupted filament is detected.

\subsubsection{During the eruption}

The eruption originates at the abovementioned abutted PIL around ($x,y$$=60$\arcsec,220\arcsec).
The event was associated to a Geostationary Operational Environmental Satellite (GOES)
B1.6 flare, which originated at coordinates N15 W04 at 09:52\,UT, peaked at 09:58\,UT,
and ended at 10:03\,UT. The full-disk H$\alpha$ images from ChroTel
exhibit two well-defined brightenings on 2016 July 3 at 09:55\,UT, next to the narrow PIL, 
one toward the north and the other one toward the south (see online movie). These are likely
the ribbons of the B-class flare. The stationarity of these ribbons indicate that the 
eruption remained confined. 
The H$\alpha$ filtergrams show bright plasma, indicating heating, 
expelled toward opposite directions (north and south), 
which about 20\,min later appears as cooler dark material that later recombines again with 
the filament channel. 
Further away from the origin of the eruption, the reason for having first bright and then 
dark H$\alpha$ features is ascribed
to the strongly wavelength-shifted spectral line during this event. When the
eruption starts, plasma is violently expelled toward higher layers of the atmosphere. 
Hence, the H$\alpha$ line is heavily blueshifted and the ChroTel filter, with a FWHM of the passband
filter between 0.5\,\AA\ and 1.0\,\AA\ \citep{bethge11}, likely only sees the wing of the line 
(higher intensities as compared to the line core). However, later 
plasma again drifts back 
toward the solar surface. Therefore, we see H$\alpha$ absorption (dark material) which apparently 
flows into the filament channel. The trajectory of the plasma therefore starts at the abutted PIL, 
where the filament eruption starts, and partially ends southward recombining with the stable 
part of the filament.

The AIA 94\,\AA, 171\,\AA, 193\,\AA, and 304\,\AA\ filtergrams nicely trace the ejected plasma, 
which appears as intensity enhancements in the data and is highlighted with a
cyan-colored arrow in Fig. \ref{Fig:AIAoverview}. 
Furthermore, the time evolution of AIA 171\,\AA\ images in Fig. \ref{Fig:AIA171} and
the animated Fig. \ref{Fig:AIAoverview} show that the slit from GREGOR was ideally 
located to record the outward moving material of the filament.  
The green arrow in Fig. \ref{Fig:AIAoverview} points to the stable part of the 
filament. The AIA 171\,\AA\ filtergrams (Fig. \ref{Fig:AIA171}) depict the evolution of
the filament eruption. At the beginning, the eruption consists of a rapid rising phase, which 
is well traced in the first three panels between 09:55\,UT and 10:05\,UT. Afterward, 
the material slows down and expands, as seen in the last three panels.

\begin{figure}[!t]
 \centering
 \includegraphics[width=0.6\hsize]{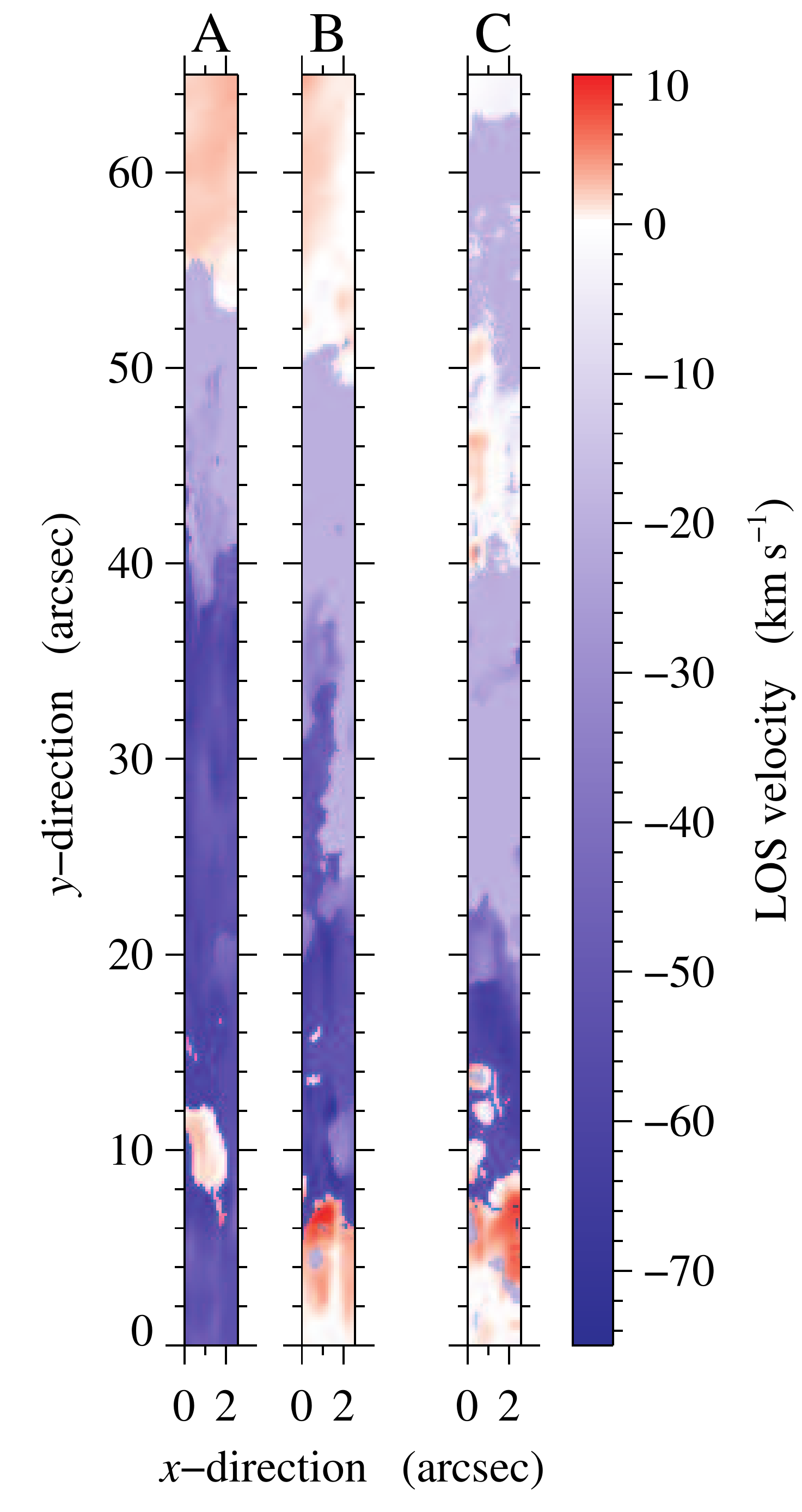} 
\caption{LOS velocity maps inferred from the inversions with HAZEL. 
From left to right, temporally consecutive maps A, B, and C are shown.
Maps A and B roughly represent the same FOV, whereas map C is 
slightly shifted toward South with respect to the first two maps. 
Up to three atmospheric components where necessary to fit the observed Stokes profiles. 
A composition of the three components is exhibited here, where always the fastest
LOS velocity is shown. 
} 
\label{Fig:vlos}
\end{figure}

\begin{figure}[!t]
 \centering
 \includegraphics[width=0.6\hsize]{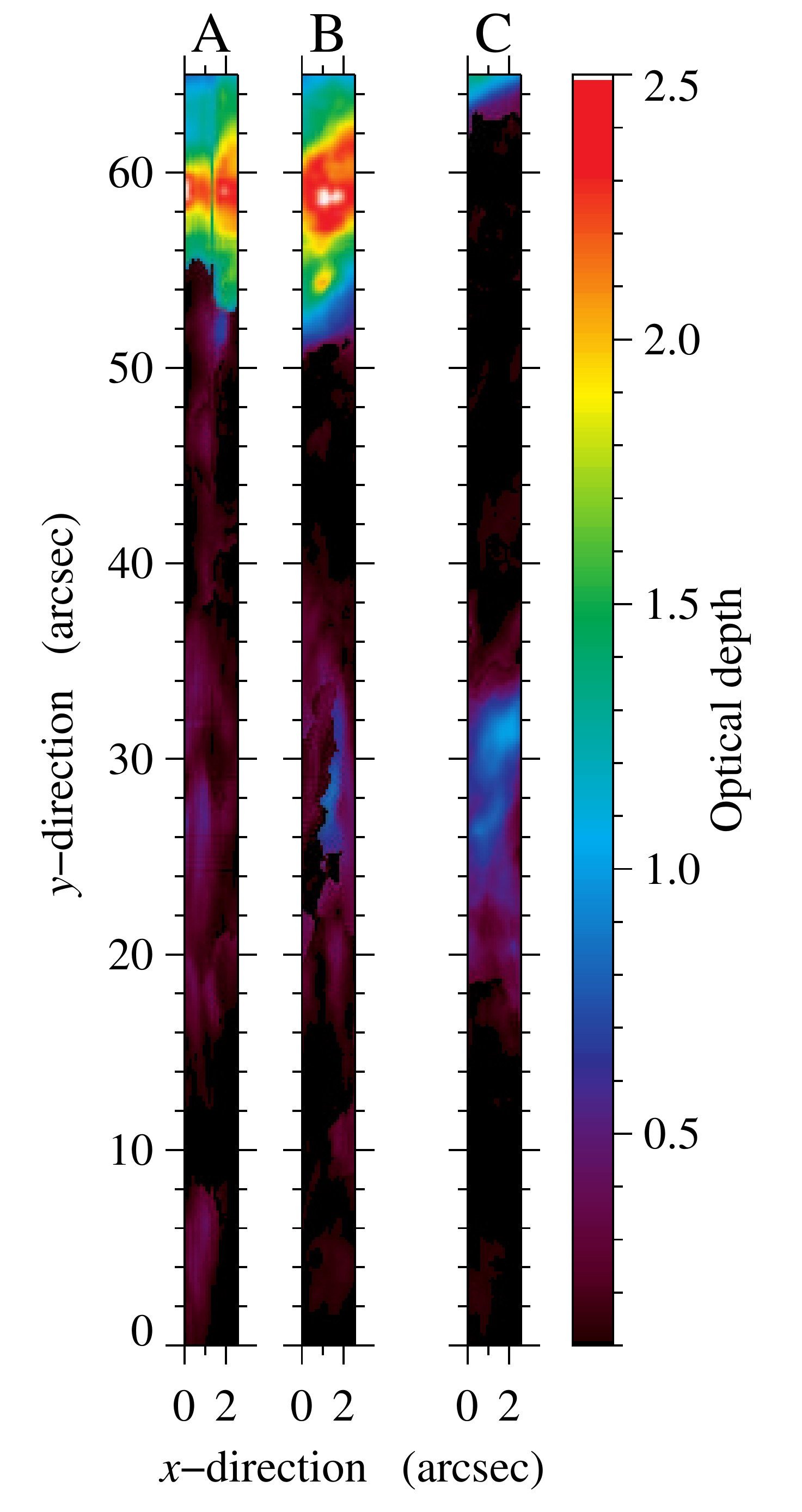} 
\caption{Same as Fig. \ref{Fig:vlos} but for the optical depth $\tau$. The maps
are a composition of up to three atmospheres. Only the optical depth corresponding 
to the fastest LOS velocity component is exhibited.}
\label{Fig:opticaldepth}
\end{figure}

\subsection{High-resolution analysis of the ejected plasma}
The slit-spectrograph GRIS at GREGOR recorded three short time series when the 
ejected plasma of the filament crossed the slit, between 10:02\,UT
and 10:10\,UT (Table \ref{tab:gris}), about 170\arcsec\ south from the 
triggering of the eruption. According to the full-disk H$\alpha$ filtergrams, the
filament eruption just began a few minutes earlier (09:55\,UT) and the expelled plasma 
was just crossing the scanned area with the slit. The top part of
the raster scans partially overlap with the spine of the original filament (best seen in 
Fig. \ref{Fig:HMIfilament}), which
remained stable. A significant amount of plasma from the eruption recombines later 
with this part of the filament.

\begin{figure}[!t]
 \centering
\includegraphics[width=0.6\hsize]{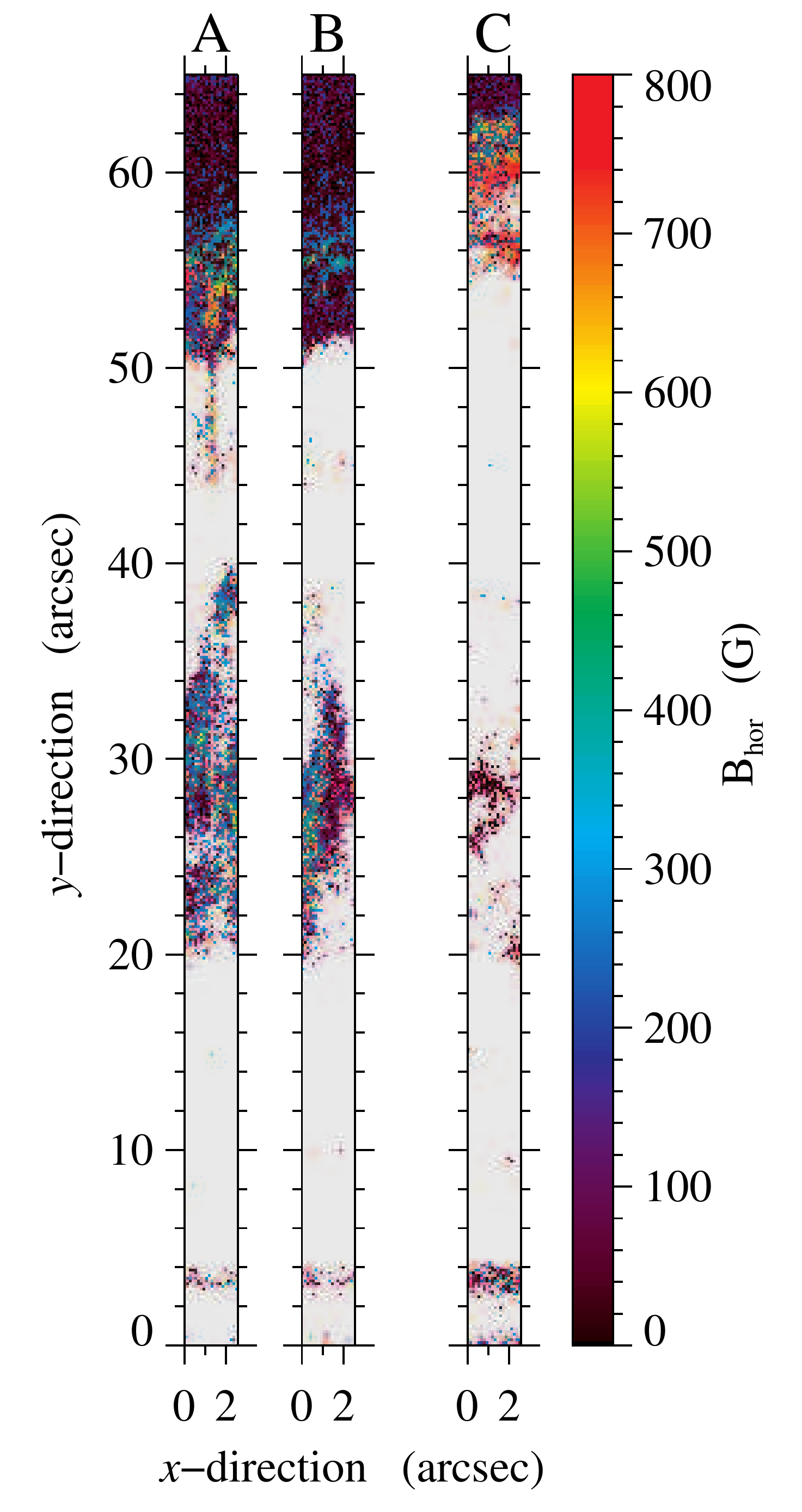} 
\caption{Same as Fig. \ref{Fig:vlos} but for the horizontal component of the magnetic field 
$B_\mathrm{hor}$. The maps are a composition of up to three atmospheres. 
Only the magnetic field corresponding to the fastest LOS velocity component is shown. 
A mask was used to exclude pixels with low polarization signals (light gray pixels). } 
\label{Fig:Bhor}
\end{figure}

\begin{figure}[!t]
 \centering
\includegraphics[width=0.6\hsize]{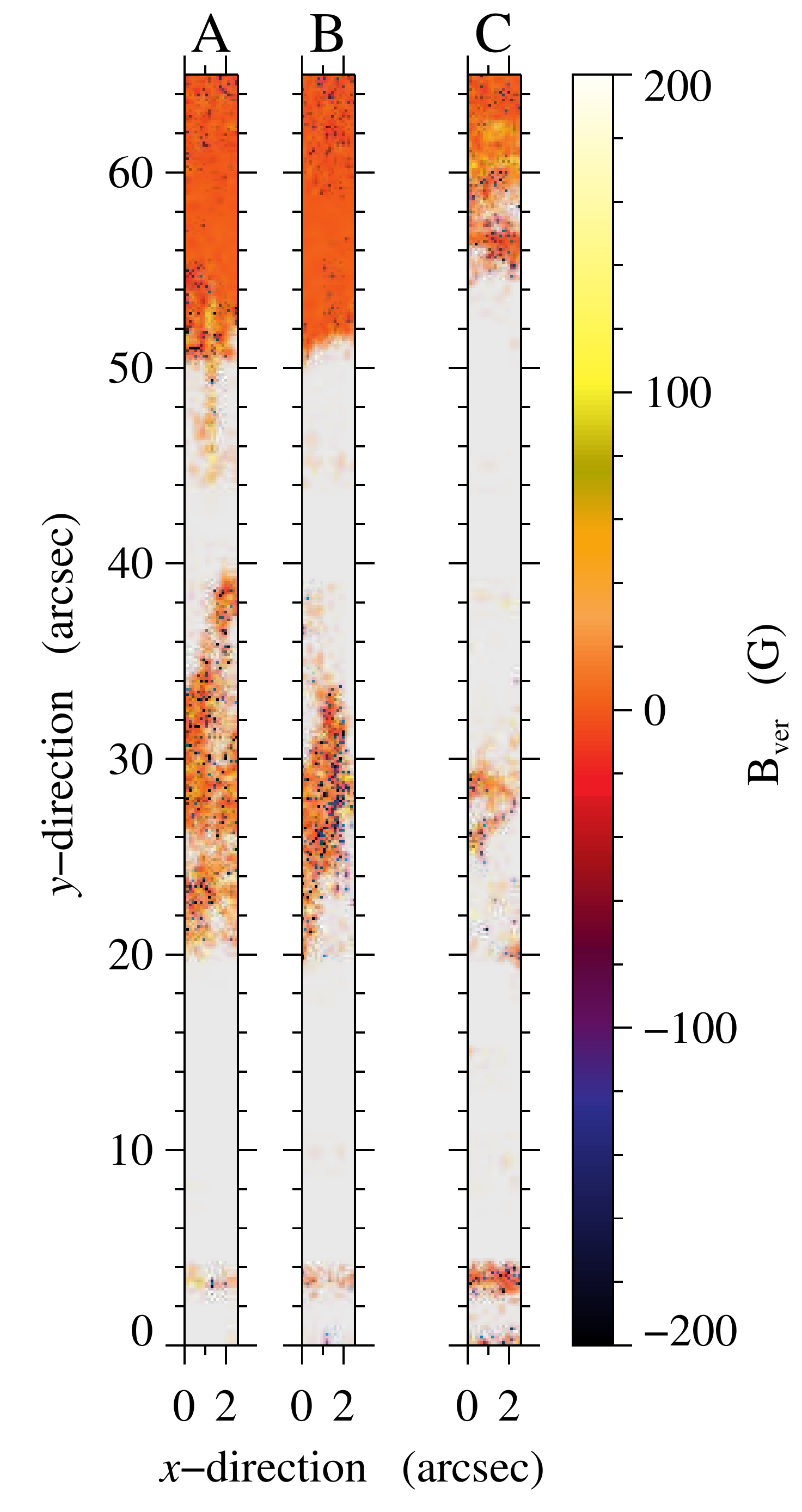} 
\caption{Same as Fig. \ref{Fig:vlos} but for the vertical magnetic field $B_\mathrm{ver}$. 
The maps are a composition of up to three atmospheres. Only the magnetic field corresponding 
to the fastest LOS velocity component is shown. A mask was used to exclude pixels with 
low polarization signals. } 
\label{Fig:Bver}
\end{figure}

The results from the inversions are illustrated in Figs. \ref{Fig:vlos}, 
\ref{Fig:opticaldepth}, \ref{Fig:Bhor}, and \ref{Fig:Bver}. 
The $y$--direction is along the slit whereas the
$x$--direction is the scan direction of the slit. Maps A and B are co-spatial, 
but map C shows a different FOV, slightly shifted to South with respect to the former maps 
(see yellow and orange slits in Figs. \ref{Fig:HMIfilament} and \ref{Fig:AIAoverview}).
Maps A--C are temporally consecutive.

\subsubsection{Line-of-sight velocity}
The LOS velocities are shown in Fig. \ref{Fig:vlos}, which displays a composition of
the three inferred atmospheres. Each pixel shows the LOS velocity arising from the
fastest atmosphere. The very fast flows are associated to the 
ejected plasma of the filament. In general, all three maps reveal strong upflows, 
with maximum values of $-$65.4\,km\,s$^{-1}$,
$-$73.0\,km\,s$^{-1}$, and $-$66.1\,km\,s$^{-1}$ in maps A--C, respectively. This is expected 
from the strong blueshifts seen of the \ion{He}{i} triplet in Fig. \ref{Fig:cluster9}. The
triplet sometimes invades significantly the red wing of the \ion{Si}{i} line.
The average LOS velocities inside the erupted filament material, between 
$y \in$ (18\arcsec, 40\arcsec), are $-$48.1\,km\,s$^{-1}$, 
$-$53.5\,km\,s$^{-1}$, and $-$42.0\,km\,s$^{-1}$, for maps A, B, and C, 
respectively. The dispersion of the velocities is high, yielding a standard 
deviation of 19.8\,km\,s$^{-1}$, 12.1\,km\,s$^{-1}$, and 26.1\,km\,s$^{-1}$, 
for maps A--C. 
Conversely, the plasma flows are slow and downward directed at the stable filament 
in the upper part of the FOV (above 52\arcsec) 
in maps A and B. We used a high optical depth ($\tau$>1.2) to isolate the pixels
belonging to the stable filament. 
The average LOS velocities within this threshold correspond to  
1.6\,km\,s$^{-1}$ and 0.5\,km\,s$^{-1}$ for maps A and B, respectively. 
The standard deviation is 0.9\,km\,s$^{-1}$ for both maps. The stable filament 
is not seen in the upper part of map C because the FOV is shifted with respect to
maps A and B.

\begin{figure}[!t]
 \centering
 \includegraphics[width=0.6\hsize]{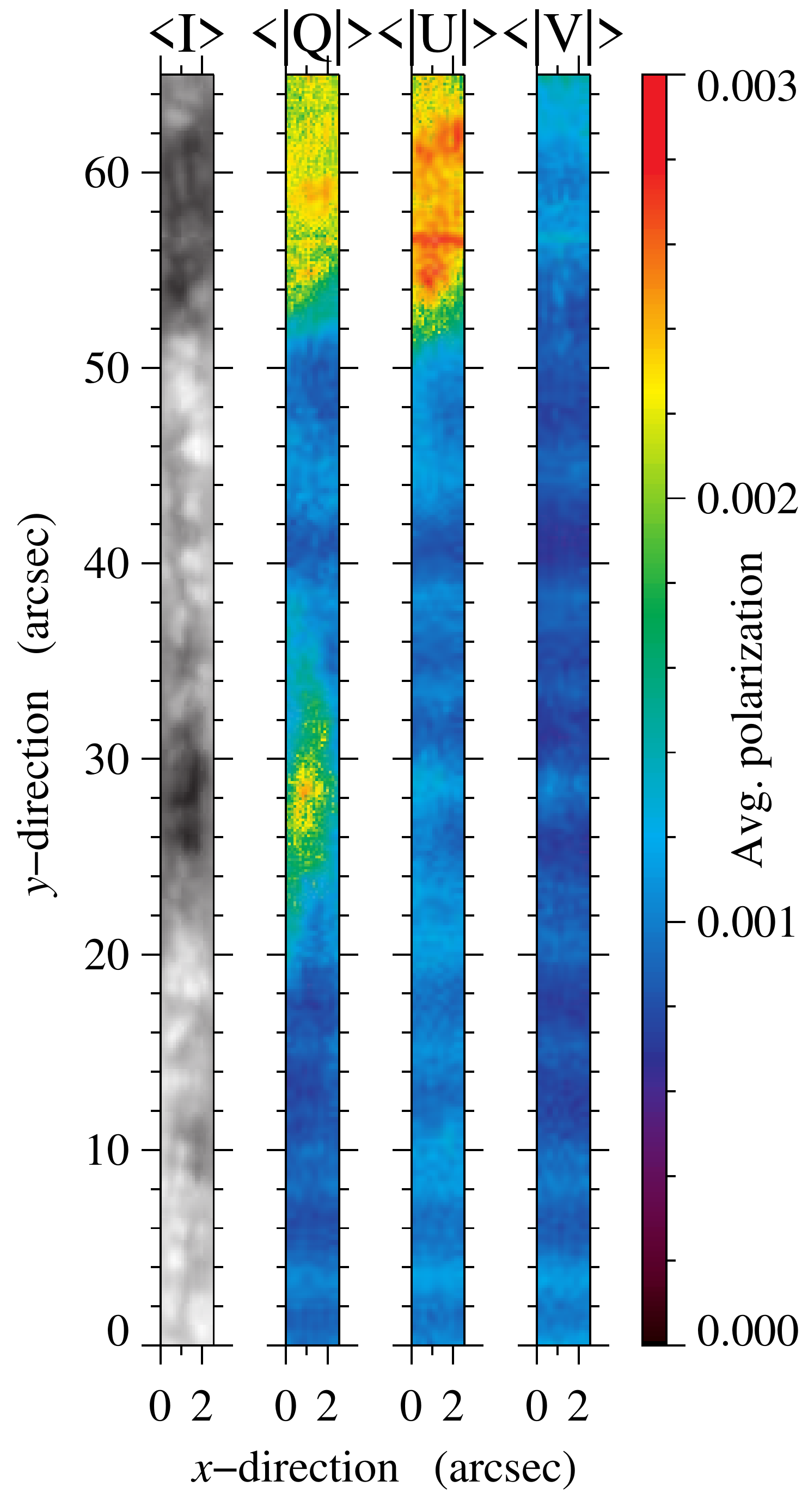} 
 \caption{Mean intensity and mean polarization signals of \ion{He}{i} 10830\,\AA\ in map B,
   between 10828.5\,\AA\ and 10831.5\,\AA. The absolute values of Stokes $|Q|/I_\mathrm{c}$,
   $|U|/I_\mathrm{c}$, and $|V|/I_\mathrm{c}$ were used and averaged to quantify the amount of      
   polarization signals.} 
 \label{Fig:pol}
\end{figure}

\subsubsection{Optical depth}
Figure \ref{Fig:opticaldepth} exhibits the optical depth of \ion{He}{i} 10830\,\AA\ 
inferred from the inversions. As mentioned above, each pixel shows the optical depth 
associated to the fastest LOS velocity atmosphere. Maps A and B
depict an increased optical depth $\tau$ above 53\arcsec\ in the $y$--axis. 
This area corresponds to the spine of the stable filament, which is at rest and remains stable 
during the partial eruption. There the optical depth reaches up to 2.5. 
An example of the strong absorption profiles, which almost reach the line depth of the
much deeper photospheric \ion{Si}{i} 10827\,\AA\ line is shown in 
Fig. \ref{Fig:cluster9} (clusters 4 and 25). 
An individual profile example from the stable filament together with its
inversion result is represented in Fig. \ref{Fig:stokes_profiles}.
The ejected plasma from the filament eruption is faintly seen as dark-violet colors all over the 
FOV (Fig. \ref{Fig:opticaldepth}). 
Between $y$$=$27\arcsec--28\arcsec, $\tau$ is slightly enhanced in map A and continues to
increase 2\,min later in map B. The optical depth in this area increases from about 0.7 to 1.1. 
Map C reaches optical-depth values up to 1.0. 

\subsubsection{Polarization and magnetic field}
\begin{figure}[!t]
 \centering
 \includegraphics[width=\hsize]{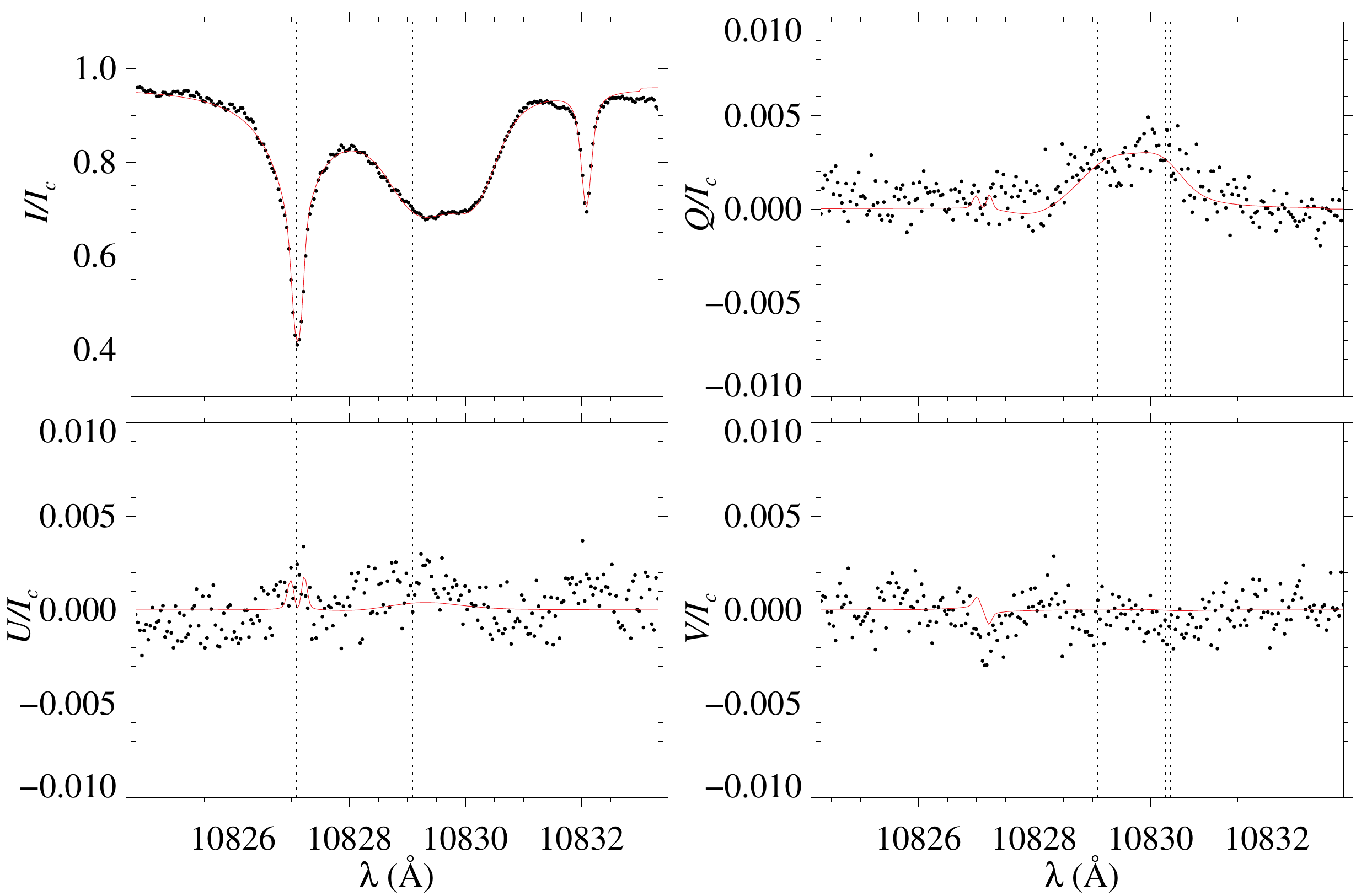} 
 \includegraphics[width=\hsize]{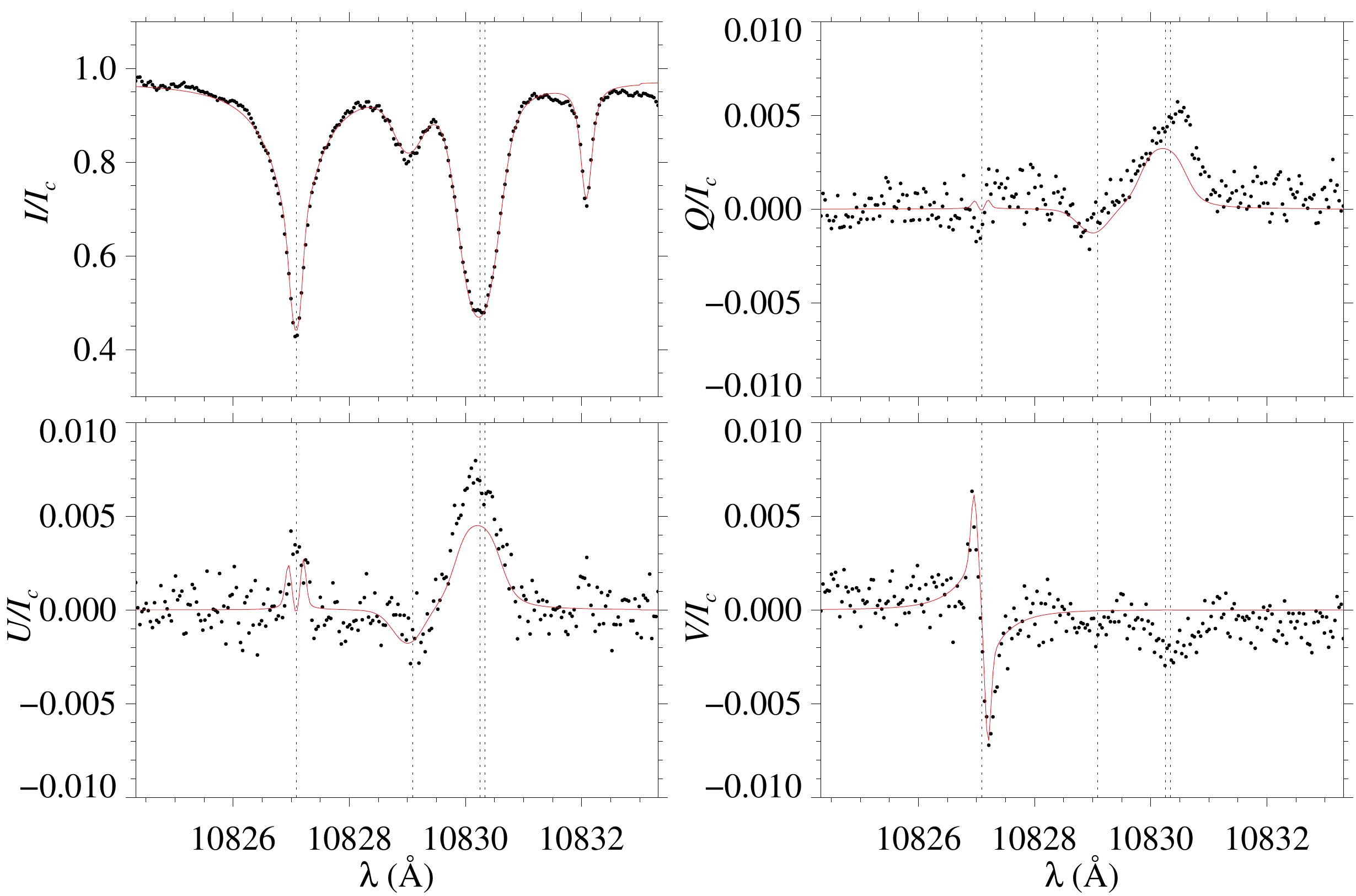} 
 \caption{Example of normalized Stokes $I/I_\mathrm{c}$, $Q/I_\mathrm{c}$, $U/I_\mathrm{c}$, 
 and $V/I_\mathrm{c}$ profiles. 
 \emph{Top}: The dots correspond to the observations from a pixel within the
 erupted filament material in map B, at coordinates ($x$,$y$) = (1\farcs7, 28\farcs4). 
 The red solid line represents the
 fit of the HAZEL inversions with two atmospheric components. The inferred horizontal magnetic 
 field is 150\,G and 23\,G, for the first and second components, respectively. 
 The vertical magnetic field is negligible and the LOS velocities are $-$4.3\,km\,s$^{-1}$
 and $-$27.8\,km\,s$^{-1}$, respectively. 
 \emph{Bottom}: the dots refer to a pixel within the filament at rest
 in the upper FOV of map B, at coordinates ($x$,$y$) = (0\farcs8, 53\farcs5). 
 The red solid line represents the
 fit of the HAZEL inversions with one atmospheric components. The inferred horizontal magnetic 
 field is 9\,G. The vertical magnetic field is negligible and the LOS velocity 
 is $-$2.1\,km\,s$^{-1}$.
 Vertical dotted lines indicate the wavelength positions at rest of \ion{Si}{i} and the \ion{He}{i}
 triplet.} 
 \label{Fig:stokes_profiles}
\end{figure}

The Stokes $Q/I_\mathrm{c}$, $U/I_\mathrm{c}$, and $V/I_\mathrm{c}$ signals within the 
three scans A, B, and C are generally small ($\sim$$10^{-3}$). 
This can be seen, for example, 
in the average polarization values in map B (Fig. \ref{Fig:pol}), which were
computed generously across the entire \ion{He}{i} triplet, between 
10828.5\,\AA\ and 10831.5\,\AA. Note that these numbers are just broad averages
to identify areas of high polarization signals. 
The largest signals arise from the stable 
filament portion in the upper part of the FOV and belong to Stokes $Q/I_\mathrm{c}$ and 
$U/I_\mathrm{c}$.  
The Stokes $Q/I_\mathrm{c}$ and $U/I_\mathrm{c}$ profiles in the filament 
(bottom four panels of Fig. \ref{Fig:stokes_profiles}) have 
a typical one-lobe Hanle effect signature, which is also expected in quiet-Sun filaments
\citep[see, e.g.,][]{trujillo02}. 
The Stokes $V/I_\mathrm{c}$
signals are negligible across the whole FOV (Fig. \ref{Fig:pol}).
Interestingly, the erupted filament material at around $y$$\sim$28\arcsec\ in Fig. \ref{Fig:pol}
only exhibits Stokes $Q$ polarization signals. We will come back to this in Sect. \ref{Sect:discussion}.

The inferred magnetic fields in Figs. \ref{Fig:Bhor} and \ref{Fig:Bver}
refer to the local solar reference frame. 
Up to three atmospheric components were needed for the \ion{He}{i} inversions. 
Here we also represent a combined map for each physical quantity. The reference for 
choosing one of the three components in each pixel is again given by the highest LOS velocity. 
We used a mask to exclude low-polarization 
signals. To that end, we computed the average polarization signal 
$\overline{P_\mathrm{QS}} = \sqrt{\bar{Q}^2+\bar{U}^2+\bar{V}^2}$ in a quiet-Sun
area, between \mbox{$x \in$ (0\arcsec--1\farcs3}) and \mbox{$y \in$ (0\arcsec--13\farcs5}), 
in each map. 
Since the polarization signals are generally small, 
we find that pixels with a $\overline{P}$ signal 
20\% above the $\overline{P_\mathrm{QS}}$, for maps A and B, represent well relevant pixels. 
However, for map C we lowered the threshold, as the degree of polarization is slightly 
lower in the entire FOV compared to maps A and B. This is likely because the erupted filament 
material is moving away. Thus, the threshold was set to
include pixels that have $\overline{P}$ signals 15\% above the $\overline{P_\mathrm{QS}}$.
The horizontal fields in Fig. \ref{Fig:Bhor} show a salt-and-pepper pattern and the results
have to be interpreted with caution. This reveals the complexity of the inversion process with
such tangled Stokes profiles. Most of these complex profiles require two or three
atmospheric components in the inversion process, which in turn leads to ambiguities and 
multiple solutions for the same observed profiles. For example, the four top panels
of Fig. \ref{Fig:stokes_profiles} show the observations (dots) and best fit (red solid line)
from the inversions. Especially Stokes $Q$ shows an extended faint lobe at the position of 
the \ion{He}{i} triplet (three rightmost vertical dashed lines), and since
Stokes $U$ is below the noise level, the result has to be interpreted with caution due to 
the arising ambiguities. Therefore, we will not rely on individual inversions and will
rather provide a statistical study of the horizontal and vertical fields inside the erupted
filament material, which is shown in Table \ref{tab:bfield}. 
For the statistics we used the results from the most blueshifted atmospheric component for each pixel.
Independent of the distribution, the $P_{50\%}$ percentile provides the median. 
The $P_{16\%}$ and $P_{84\%}$ percentiles are shown to indicate the equivalent range of $\pm$1 
standard deviation $\sigma$ under the assumption of a normal distribution. 
Maps A and B show on average stronger horizontal fields $\overline{B_\mathrm{hor}}$ of 
254\,G and 262\,G, respectively, than map C, which yields on average 173\,G, but in the latter map the area of the erupted material is 
much smaller.

\begin{table}[!ht]
\begin{center}
\caption{Horizontal and vertical magnetic field statistics. 
The average field strength and three 
different percentiles $P$, corresponding to the cumulative percentages 
of $-1\sigma$, median, and $+1\sigma$
of a normal distribution, are shown. The number of pixels belonging to the ejected plasma for 
maps A, B, and C are: 1635, 1119, and 285.}\label{tab:bfield}
\begin{tabular}{ccccc}
\hline
\hline
Map             & $\overline{B_\mathrm{hor}}\, (G)$ & $P_{16\%}$  & $P_{50\%}$  & $P_{84\%}$  \rule[-4pt]{0pt}{15pt}  \\
\hline
\hline
A               &   254  & 53  & 207  &  467    \rule[-4pt]{0pt}{14pt} \\
B               &   262  & 55  & 165  &  544     \\
C               &   173  &  4  &  27  &  560     \\
\hline
             & $\overline{|B_\mathrm{ver}|}$\, (G) & $P_{16\%}$  & $P_{50\%}$  & $P_{84\%}$  \rule[-4pt]{0pt}{15pt} \\
\hline
\hline
A               &   58  &  1 &  20  &  130  \rule[-4pt]{0pt}{14pt} \\
B               &   78  &  2 &  45  &  164   \\
C               &   39  &  3 &  22  &   77   \\
\hline
\end{tabular}
\end{center}
\end{table}

The vertical magnetic fields are smaller than the horizontal ones, which is not surprising, 
as the Stokes $V$ signals are almost absent (Fig. \ref{Fig:pol}). We find on average a 
$\overline{|B_\mathrm{ver}|}$ of 
58\,G and 78\,G for maps A and B, respectively, within the area of the erupted filament, 
and 39\,G for map C (Table \ref{tab:bfield}). 

Regarding the area of the stable filament, that is, above $y$$=$53\arcsec\ in maps A and B, 
we find as well salt-and-pepper-like horizontal fields. However, the Stokes $Q$ and $U$ 
signals are significantly different from the Stokes profiles found in the erupted material. 
Fig. \ref{Fig:stokes_profiles}
compares one example of Stokes profiles of the stable filament with the erupting material, 
together with its best inversion fit.  
The linear polarization profiles, that is, Stokes $Q$ and $U$, are larger and tighter than in the 
erupted material. Stokes $I$ shows a deep and unshifted \ion{He}{i} triplet. 
The vertical fields in the filament are almost absent (Fig. \ref{Fig:Bver}), as 
expected in the spine of the filament. This can be seen in the lack of Stokes $V$
in the bottom panel of Fig. \ref{Fig:stokes_profiles}. 

\begin{figure}[!t]
 \centering
 \includegraphics[width=0.6\hsize]{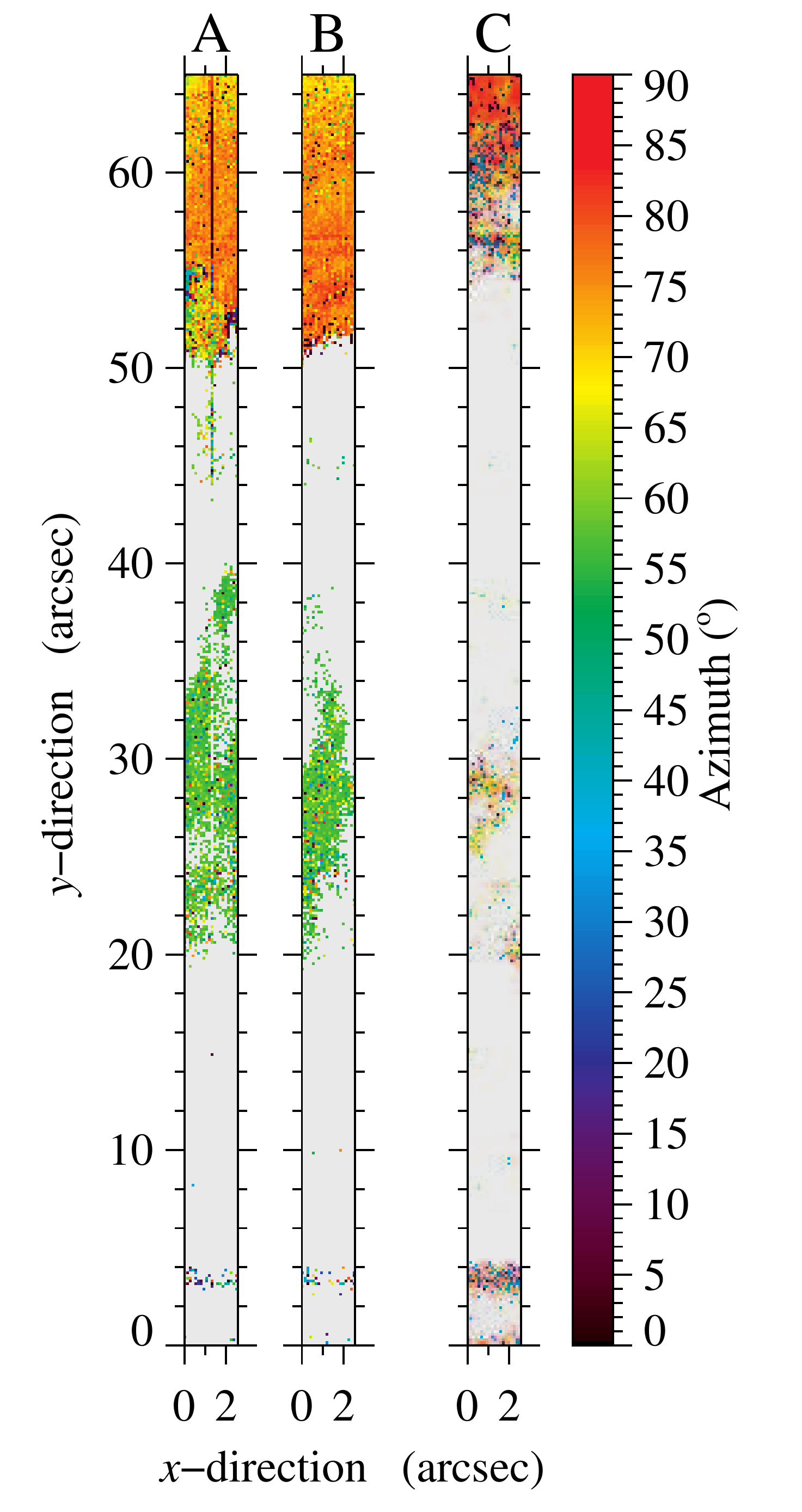} 
\caption{Same as Fig. \ref{Fig:vlos} but for the azimuth of the magnetic field $\phi$. 
One solution, corresponding to the first quadrant $\phi \in [0^\circ,90^\circ]$,
is shown. } 
\label{Fig:azimuth}
\end{figure}

We compute the azimuth $\phi$ of the magnetic field lines using
\begin{equation} \label{eq:1}
    \phi = \arctan \left( \frac{B_y}{B_x} \right) ~, 
\end{equation}
where $B_y$ and $B_x$ are the inferred horizontal components of the magnetic field from the inversions. 
Due to the potential ambiguities which arise from the inference of the magnetic field using the Stokes $Q$, $U$, 
and $V$ profiles, we have different possible solutions for the azimuth
($\pm 180^\circ$ and $\pm 90^\circ$ in
the azimuth in the LOS reference frame). By assuming that the
field lines have an homogeneous distribution, rather than an unorganized one, we find a smooth solution
for the azimuth by rotating all angles to the first quadrant (Q$_1$ where $\phi \in [0^\circ,90^\circ]$) of
a Cartesian two-dimensional system. This rotation was achieved by showing
\mbox{$\phi \mod 90^\circ$}. We use the fact that the observations are
very close to disk center, so that the inferred azimuth is very similar
to the azimuth in the LOS. The results are shown in Fig. \ref{Fig:azimuth}. The average azimuth 
$\overline{\phi}$ and its standard 
deviation $\sigma$ appear in Table \ref{tab:azimuth}. The other possible solutions for all
quadrants are also mentioned in the table. Any combinations between quadrants Q$_1$--Q$_4$ 
of the erupted filament and the stable filament are possible. Maps A and B show almost identical 
numbers (within decimals), therefore we merged both results in the table. 
Unfortunately map C does not show the stable filament in the
upper part of the slit  owing to its shifted FOV (Fig. \ref{Fig:opticaldepth}).  

\begin{table}[!ht]
\begin{center}
\caption{Possible average azimuth $\overline{\phi}$ configurations for the erupted 
and stable filament areas within
maps A and B. Any combinations between quadrants Q$_1$--Q$_4$ of the erupted filament and the 
Q$_1$--Q$_4$ of the stable filament are possible. The dispersion associated to
the average azimuth is given by the standard deviation $\sigma$. }\label{tab:azimuth}
\begin{tabular}{ccccccccc}
\hline
\hline
                & \multicolumn{8}{c}{Azimuth $\overline{\phi}$ ($^\circ$)} \rule[-4pt]{0pt}{15pt}  \\
                & \multicolumn{4}{c}{Erupted filament} &  \multicolumn{4}{c}{Stable filament} \\
Maps            & Q$_1$ & Q$_2$ & Q$_3$ & Q$_4$ & Q$_1$ & Q$_2$ & Q$_3$ & Q$_4$ \\
\hline
\hline
A \& B          &   60  &  150   & 240  & 330  &  78  & 168  & 258  & 348    \rule[-4pt]{0pt}{14pt} \\
$\sigma$        & \multicolumn{4}{c}{11} &  \multicolumn{4}{c}{14}  \\
\hline
\end{tabular}
\end{center}
\end{table}

%--------------------------------------------------------------------------
\section{Discussion} \label{Sect:discussion}
%--------------------------------------------------------------------------
Our high spatial- and spectral-resolution data from the infrared spectrograph
attached to the 1.5-meter GREGOR telescope show intriguing observations of the material
of a filament eruption. The triggering of the eruption itself was not covered by
the high-resolution observations. However, full-disk H$\alpha$ images (Fig. \ref{Fig:Chrotel})
and AIA/SDO images (Fig. \ref{Fig:AIAoverview}) reveal that the origin was located next to an 
abutted PIL and was associated to a B1.6 flare.
A segment of the filament then rose, moved southward, and crossed our FOV about ten minutes later, 
while the remaining filament remained stable. The ejected plasma is well seen in the hotter 
AIA channels, for instance, 94\,\AA, 171\,\AA\, and 193\,\AA, and in addition 
in the chromospheric \ion{He}{ii} 304\,\AA\ filtergrams 
(Figs. \ref{Fig:AIAoverview} and \ref{Fig:AIA171}).
Fragmentary eruptions are not uncommon
and often partially expelled material recombines with the original filament
\citep[see, e.g.,][]{jenkins18, yan20}. This is also happening in our case and is best seen in the context 
H$\alpha$ full disk images. 

\subsection{Variety of spectral profiles}
A large variety of spectral profiles were found in the observations, which demonstrates the
complexity of analyzing very dynamic and eruptive events on the Sun. \citet{sasso11} and
\citet{schad16} presented strongly redshifted Stokes profiles in their 
\ion{He}{i} 10830\,\AA\ observations, with LOS
velocities of up to 100\,km\,s$^{-1}$ and 185\,km\,s$^{-1}$, respectively. Both works
used the inversion code He-Line Information Extractor \citep[HELIX,][]{lagg04} with several
atmospheric components. However, we took a different inversion approach. We first classified
and put into groups all similar Stokes $I$ profiles across the FOV using the 
unsupervised machine-learning 
algorithm $k$-means. We came up with 30 representative clusters (Fig. \ref{App:A}), which 
cover the spectral variety. Then, according to the shape of the average intensity profile
of each group, we chose up to three atmospheric components, each one having a different initial 
set of velocity 
ranges. This was crucial for the successful convergence of the inversion code 
because the initial guess atmospheres
comprised velocity ranges which were already close to the observed ones. In addition, 
it allowed us to disentangle the different features within the observed FOV.
There were clearly two important groups of
spectral profiles inside the FOV (Fig. \ref{Fig:cluster9}): (1) the erupted material 
(for example, clusters 2 and 10) and
(2) the filament at rest (clusters 4 and 25). An example of the four Stokes parameters of 
each group is shown in Fig. \ref{Fig:stokes_profiles}.
The first group illustrates similar profiles as presented by
\citet{sasso11}. However, their profiles exhibit a deeper absorption of \ion{He}{i}, 
because they were optically thicker than ours. 
Surprisingly, compared to the filament eruption shown by 
\citep{sasso11, sasso14}, there are no downflows in our erupted plasma. The authors
showed Stokes $I$ profiles which have both, blueshifted and redshifted \ion{He}{i} absorption lines. 
Since we do not detect that in our observations, a plausible explanation is that our captured 
material must have belonged to a strongly-rising phase of the eruption process. 
The H$\alpha$ and SDO movies show that the rising filament material partially falls 
back to the filament channel. Unfortunately, these times
were not covered by our GREGOR observations.  Strong redshifted \ion{He}{i} profiles indicating draining of the 
plasma may be expected there. 

\subsection{High velocities of the erupted material}
We analyzed three consecutive maps (Table \ref{tab:gris}), taken in a short time scale (<\,8\,min). 
Maps A and B are roughly co-spatial and they cover about four minutes of the 
event. Within this short time scale we see significant changes in the optical depth of the
erupted filament material (Fig. \ref{Fig:opticaldepth}). 
However, the inferred magnetic fields display very similar values (Figs. \ref{Fig:Bhor} and 
\ref{Fig:Bver}). The changes in the shape 
of the \ion{He}{i} cloud are ascribed to the fast crossing of the material throughout our FOV.
The velocities reported for filament eruptions are widespread. It is often distinguished 
between a slow rising phase (for example, 10--15\,km\,s$^{-1}$) followed by a rapid ejection of the material
(for example, 100--200\,km\,s$^{-1}$) \citep{sterling04,sterling05,penn00, dhara17}. 
The velocities also depend on the height up to which the filament eruption can be traced, for example,  
600\,km\,s$^{-1}$ were found in transition-region spectral-line Doppler shifts \citep{kleint15}, 
and even higher velocities when tracing CME material \citep[see, e.g.,][]{moon02, cheng20}.

At the height of \ion{He}{i} 10830\,\AA\ formation we find LOS velocities of up to 
$\sim$73\,km\,s$^{-1}$
(Fig. \ref{Fig:vlos}), which are upflows given the proximity to the disk center.
We interpret this value as a lower limit, since the \ion{He}{i} triplet runs into the 
neighboring \ion{Si}{i} line and cannot be reliably distinguished anymore by the inversion code. 
The largest amount of erupted material is seen in map A, followed by map B, while map C shows 
the smallest amount (Fig. \ref{Fig:opticaldepth}). This is consistent with a cloud of \ion{He}{i}
rapidly moving horizontally from left to right across the FOV. The AIA/SDO filtergrams confirm
this motion of the plasma, which we roughly estimate at $\sim$100\,km\,s$^{-1}$, from the 
tracking of bright features of the filament in SDO/AIA images. This motion is  
roughly perpendicular to the slit orientation in the movie. 
Combining the fastest LOS velocities inferred from our \ion{He}{i} observations
with the estimated projected velocities from SDO/AIA we compute the total velocity
as  
\begin{equation}
    v = \sqrt{v_\mathrm{He}^2+v_\mathrm{AIA}^2} ~~,
\end{equation} 
which yields a lower limit of $v$$\sim$124\,km\,s$^{-1}$ for the erupted 
plasma of our filament. 

\subsection{Magnetic field configuration}
The erupted material shows faint but coherent polarization signals,
which result from the Hanle effect, indicating an organized orientation 
of the magnetic field lines (Fig. \ref{Fig:pol}). Moreover, there is no indication 
of circular polarization or Zeeman-effect patterns. 
Only Stokes $Q$ is found in the area of high \ion{He}{i}
absorption belonging to the erupted material of the quiescent filament. Conversely, inside the
stable filament, Stokes $Q$ and $U$ is prominently present, indicating that a different
orientation of the magnetic field lines exists. 
We quantified this orientation by computing the average azimuth of the magnetic field lines, 
obtained with Eq. \ref{eq:1}, in the erupted and in the stable filament (Table \ref{tab:azimuth}). 
When rotating the azimuth angles to the first quadrant Q$_1$, we discover a smooth 
solution (Fig. \ref{Fig:azimuth}) as expected from the homogeneous polarization signals. 
However, this solution is not unique. To search for differences between the magnetic structure of the
erupting material and the stable filament, we rotate the azimuth solution by multiples of 
90$^\circ$. The outcome is that there is no
possible combination of the several quadrants that satisfy that the stable filament is aligned 
with the erupted plasma of the filament. 
This demonstrates that both magnetic systems have a different orientation of the field lines.
The closest possible orientation appears if both, the erupted material and the stable filament, 
lie in the 
first quadrant Q$_1$. Then the difference of the azimuth angle between both structures is only
$\sim$18$^\circ$.

We cannot provide the absolute orientation of the magnetic field lines with respect 
to the solar surface. The reason are the intrinsic ambiguities bound to the Stokes profiles themselves.
In addition, more ambiguities arise from the use of up to three atmospheric components, 
which are necessary to fit the observed Stokes profiles. The use of multiple components was already 
reported by \citet{sasso11} and was ascribed to different layers along the LOS, 
highlighting the importance of the fine structure in filaments. 
When combining the results of our three atmospheric components -- taking into account only the
most blueshifted atmosphere, as it represents best the erupted 
plasma -- we find predominantly horizontal magnetic fields of
254\,G and 262\,G, (Table \ref{tab:bfield}), in the first two maps, respectively. 
The average vertical component of the magnetic field 
in the erupted material for these two maps is lower, 58\,G and 78\,G, respectively. 
Hence, the total field strength lies between typical values inferred for quiescent
filaments \citep[20--40\,G, e.g.,][]{trujillo02, merenda06} and AR filaments 
\citep[600--800\,G,][]{kuckein09, guo10, xu12}. The retrieved number is more than twice higher than the 
average total magnetic field of $\sim$119\,G inferred by \citet{sasso14}, which the 
authors found in their \ion{He}{i} absorption cloud associated to their active region filament eruption.
Note that our filament does not belong to an active region and therefore shows 
uncommonly high field strengths compared to stable quiescent filaments.
Whether the strong fields might be a consequence of an ambiguity cannot be ruled out 
at this point, since no other chromospheric lines with polarimetry are available.  

Although the erupted plasma has an organized orientation (Fig. \ref{Fig:pol} and movie of 
Fig. \ref{Fig:AIAoverview}), 
from the analyzed data we cannot trace individual field lines. 
\citet{wang20} interpreted in their observations a flux rope topology 
for the filament, which does not change during the 
initial phase of their eruption. Furthermore, \citet{xue16} reported on untwisting of the flux rope 
in an erupted filament. Our inferred field strengths and observed polarization maps
are not incompatible with such a flux rope topology. 
However, we do not find evidences for rotating
motions of the erupting plasma, as described by \citet{li17}. If there was such a rotation,
we would find opposite directed flows on both sides close to the main axis of the flux rope 
of the erupted material. 
According to the SDO/AIA (Fig. \ref{Fig:AIAoverview}) and the full-disk H$\alpha$ 
filtergrams, the main axis of the material should be approximately perpendicular to our $y-$axis
(along the slit), in maps A and B. 
The redshifts, which are found in our observations, are outside of the erupted
material, represented by higher optical depth (Fig. \ref{Fig:opticaldepth}).
In addition, the average 
azimuth angle of the magnetic field between the two consecutive maps A and B is virtually the same.
Both maps were observed within 4\,min and 13\,s. In this time range no 
changes in the field orientation are seen. 
The AIA 171\,\AA\ filtergrams in Fig. \ref{Fig:AIA171} 
(see also movie of Fig. \ref{Fig:AIAoverview}), 
suggest untwisting of the erupted filament in the time range 10:10--10:40\,UT. 
Hence, this happens after our high-resolution observations indicating
that we just observed the fast rising phase of the filament. 
The untwisting stage was observed here only about 15\,min
after the start of the filament eruption, which we pinpointed at 09:55\,UT according to the 
full-disk H$\alpha$ images.

%--------------------------------------------------------------------------
\section{Conclusions}
%--------------------------------------------------------------------------
We characterized the plasma of an erupting filament using full-Stokes spectral-line
inversions. We found fast moving plasma with a total velocity of $\sim$124 \,km\,s$^{-1}$,
as a lower limit. Furthermore, 
predominantly horizontal magnetic fields, on average between 173--262\,G, were found, 
which is untypically high for quiescent filaments.
The field lines were smoothly organized, showing a low dispersion, of the order of 11$^\circ$, in the
azimuth angle. The orientation remained stable within two consecutive
observed maps. This indicates no rotation of the field lines at 
fast time scales below $\sim$2.5 minutes, during the fast-eruption phase. 
In our case, the untwisting phase started about 15\,min after the filament eruption started.

Erupting-filament observations with polarimetry from ground-based high-resolution telescopes are
rare and, to our knowledge, this is the first filament eruption observed by GREGOR.
Difficulties arise mainly due to the unpredictability of such spontaneous
events. The present data have shown the complexity of Stokes profiles associated to
eruptive solar events such as filament eruptions. 
There are no current inversion codes to automatically deal with the 
numerous ambiguities arising from the observed polarization signals. One important 
step to disentangle the various signals is using machine-learning algorithms such as 
the $k$-means used in this work. In the future, a combination of such classification
algorithms together with complex inversion codes that provide also the probability 
of the possible solutions are necessary. 
The use of extensive multiwavelength polarimetric observations,
as planned, for example, for the Daniel K. Inouye Solar Telescope \citep[DKIST;][]{DKIST} and
the European Solar Telescope \citep[EST;][]{EST}, will also significantly reduce the number of 
ambiguities.

% =====================================================================
\begin{acknowledgements}
The 1.5-meter GREGOR solar telescope was built by a German consortium under the 
leadership of the Leibniz-Institut f\"ur Sonnenphysik in Freiburg (KIS) with the
Leibniz-Institut f\"ur Astrophysik Potsdam (AIP), 
the Institut f\"ur Astrophysik G\"ottingen (IAG), the Max-Planck-Institut f\"ur
Sonnensystemforschung in G\"ottingen (MPS), and the Instituto de Astrof\'isica 
de Canarias (IAC), and with contributions by the Astronomical 
Institute of the Academy of Sciences of the Czech Republic (ASCR).
The ChroTel filtergraph has been developed by the KIS in cooperation with the High 
Altitude Observatory (HAO) in Boulder, Colorado, USA. 
We thank B. Kliem for helpful comments on the manuscript.
We thank the 
referee for providing helpful suggestions to improve the manuscript.
The observations were supported by the SOLARNET access time, which belonged 
to the European Commission’s 7th Framework Programme under grant agreement No. 312495.
CK and SJGM thank the German Academic Exchange Service (DAAD) for 
funds from the German Federal Ministry of Education \& Research and the
Slovak Academy of Science under project No. 57449420. CK acknowledges KIS for 
travel support within this collaboration. 
Funding from the Horizon 2020 projects SOLARNET (No 824135) and ESCAPE (No 824064)
is gratefully acknowledged. SJGM acknowledges the support of the project 
VEGA 2/0048/20. SJGM also is grateful for the support of the Stefan 
Schwarz grant of the Slovak Academy of Sciences and the support by 
the Erasmus+ programme of the European Union under grant 
number 2017-1-CZ01-KA203-035562 during his 2019 stay at the IAC. 
SJGM was partially 
supported by the Spanish Ministry of Science, Innovation and Universities through the 
grant PGC2018- 095832-B-I00, and by the European Research
Council through the grant ERC- 2017-CoG771310-PI2FA. 
AAR acknowledges financial
support from the Spanish Ministerio de Ciencia, Innovaci\'on y Universidades
through project PGC2018-102108-B-I00 and FEDER funds.
This research has made use of NASA's Astrophysics Data System.

\end{acknowledgements}

%===============================================================================
%    BIBLIOGRAPHY
%===============================================================================

\bibliographystyle{aa}
\bibliography{aa-jour,biblio}

\begin{thebibliography}{70}
\expandafter\ifx\csname natexlab\endcsname\relax\def\natexlab#1{#1}\fi

\bibitem[{{Amari} {et~al.}(1999){Amari}, {Luciani}, {Mikic}, \&
  {Linker}}]{amari99}
{Amari}, T., {Luciani}, J.~F., {Mikic}, Z., \& {Linker}, J. 1999, \apjl, 518,
  L57

\bibitem[{{Antiochos} {et~al.}(1999){Antiochos}, {DeVore}, \&
  {Klimchuk}}]{antiochos99}
{Antiochos}, S.~K., {DeVore}, C.~R., \& {Klimchuk}, J.~A. 1999, \apj, 510, 485

\bibitem[{{Asensio Ramos} {et~al.}(2008){Asensio Ramos}, {Trujillo Bueno}, \&
  {Land i Degl'Innocenti}}]{hazel08}
{Asensio Ramos}, A., {Trujillo Bueno}, J., \& {Land i Degl'Innocenti}, E. 2008,
  \apj, 683, 542

\bibitem[{{Avrett} {et~al.}(1994){Avrett}, {Fontenla}, \& {Loeser}}]{avrett94}
{Avrett}, E., {Fontenla}, J., \& {Loeser}, R. 1994, in IAU Symposium, Vol. 154,
  Infrared Solar Physics, ed. D.~M. {Rabin}, J.~T. {Jefferies}, \&
  C.~{Lindsey}, 35

\bibitem[{{Babcock} \& {Babcock}(1955)}]{babcock55}
{Babcock}, H.~W. \& {Babcock}, H.~D. 1955, \apj, 121, 349

\bibitem[{{Berkefeld} {et~al.}(2010){Berkefeld}, {Soltau}, {Schmidt}, \& {von
  der L{\"u}he}}]{Berkefeld10}
{Berkefeld}, T., {Soltau}, D., {Schmidt}, D., \& {von der L{\"u}he}, O. 2010,
  {Appl. Opt.}, 49, G155

\bibitem[{{Bethge} {et~al.}(2011){Bethge}, {Peter}, {Kentischer},
  {Halbgewachs}, {Elmore}, \& {Beck}}]{bethge11}
{Bethge}, C., {Peter}, H., {Kentischer}, T.~J., {et~al.} 2011, \aap, 534, A105

\bibitem[{{Canou} \& {Amari}(2010)}]{canou10}
{Canou}, A. \& {Amari}, T. 2010, \apj, 715, 1566

\bibitem[{{Cheng} {et~al.}(2020){Cheng}, {Zhang}, {Kliem}, {\{T{\"o}r{\"o}k\}},
  {Xing}, {Zhou}, {Inhester}, \& {Ding}}]{cheng20}
{Cheng}, X., {Zhang}, J., {Kliem}, B., {et~al.} 2020, arXiv e-prints,
  arXiv:2004.03790

\bibitem[{{Collados}(1999)}]{collados99}
{Collados}, M. 1999, in ASP Conf.\ Ser., Vol. 184, Third Advances in Solar
  Physics Euroconference: Magnetic Fields and Oscillations, ed. {B.~Schmieder,
  A.~Hofmann, \& J.~Staude}, 3--22

\bibitem[{{Collados} {et~al.}(2012){Collados}, {L{\'o}pez}, {P{\'a}ez},
  {Hern{\'a}ndez}, {Reyes}, {Calcines}, {Ballesteros}, {D{\'{\i}}az}, {Denker},
  {Lagg}, {Schlichenmaier}, {Schmidt}, {Solanki}, {Strassmeier}, {von der
  L{\"u}he}, \& {Volkmer}}]{collados12}
{Collados}, M., {L{\'o}pez}, R., {P{\'a}ez}, E., {et~al.} 2012, AN, 333, 872

\bibitem[{{Collados}(2003)}]{collados03}
{Collados}, M.~V. 2003, in Society of Photo-Optical Instrumentation Engineers
  (SPIE) Conference Series, ed. {S.~Fineschi}, Vol. 4843, 55--65

\bibitem[{{Dhara} {et~al.}(2017){Dhara}, {Belur}, {Kumar}, {Banyal}, {Mathew},
  \& {Joshi}}]{dhara17}
{Dhara}, S.~K., {Belur}, R., {Kumar}, P., {et~al.} 2017, \solphys, 292, 145

\bibitem[{{D{\'\i}az Baso} {et~al.}(2016){D{\'\i}az Baso}, {Mart{\'\i}nez
  Gonz{\'a}lez}, \& {Asensio Ramos}}]{diazbaso16}
{D{\'\i}az Baso}, C.~J., {Mart{\'\i}nez Gonz{\'a}lez}, M.~J., \& {Asensio
  Ramos}, A. 2016, \apj, 822, 50

\bibitem[{{D{\'\i}az Baso} {et~al.}(2019{\natexlab{a}}){D{\'\i}az Baso},
  {Mart{\'\i}nez Gonz{\'a}lez}, \& {Asensio Ramos}}]{diaz_baso19a}
{D{\'\i}az Baso}, C.~J., {Mart{\'\i}nez Gonz{\'a}lez}, M.~J., \& {Asensio
  Ramos}, A. 2019{\natexlab{a}}, \aap, 625, A128

\bibitem[{{D{\'\i}az Baso} {et~al.}(2019{\natexlab{b}}){D{\'\i}az Baso},
  {Mart{\'\i}nez Gonz{\'a}lez}, \& {Asensio Ramos}}]{diaz_baso19b}
{D{\'\i}az Baso}, C.~J., {Mart{\'\i}nez Gonz{\'a}lez}, M.~J., \& {Asensio
  Ramos}, A. 2019{\natexlab{b}}, \aap, 625, A129

\bibitem[{{Doyle} {et~al.}(2019){Doyle}, {Wyper}, {Scullion}, {McLaughlin},
  {Ramsay}, \& {Doyle}}]{doyle19}
{Doyle}, L., {Wyper}, P.~F., {Scullion}, E., {et~al.} 2019, \apj, 887, 246

\bibitem[{{Gonz{\'a}lez Manrique} {et~al.}(2016){Gonz{\'a}lez Manrique},
  {Kuckein}, {Pastor Yabar}, {Collados}, {Denker}, {Fischer}, {G{\"o}m{\"o}ry},
  {Diercke}, {Bello Gonz{\'a}lez}, {Schlichenmaier}, {Balthasar}, {Berkefeld},
  {Feller}, {Hoch}, {Hofmann}, {Kneer}, {Lagg}, {Nicklas}, {Orozco Su{\'a}rez},
  {Schmidt}, {Schmidt}, {Sigwarth}, {Sobotka}, {Solanki}, {Soltau}, {Staude},
  {Strassmeier}, {Verma}, {Volkmer}, {von der L{\"u}he}, \&
  {Waldmann}}]{gonzalez16}
{Gonz{\'a}lez Manrique}, S.~J., {Kuckein}, C., {Pastor Yabar}, A., {et~al.}
  2016, Astronomische Nachrichten, 337, 1057

\bibitem[{{Guo} {et~al.}(2010){Guo}, {Schmieder}, {D{\'e}moulin}, {Wiegelmann},
  {Aulanier}, {T{\"o}r{\"o}k}, \& {Bommier}}]{guo10}
{Guo}, Y., {Schmieder}, B., {D{\'e}moulin}, P., {et~al.} 2010, \apj, 714, 343

\bibitem[{{Hofmann} {et~al.}(2012){Hofmann}, {Arlt}, {Balthasar}, {Bauer},
  {Bittner}, {Paschke}, {Popow}, {Rendtel}, {Soltau}, \&
  {Waldmann}}]{hofmann12}
{Hofmann}, A., {Arlt}, K., {Balthasar}, H., {et~al.} 2012, AN, 333, 854

\bibitem[{{Jenkins} {et~al.}(2018){Jenkins}, {Long}, {van Driel-Gesztelyi}, \&
  {Carlyle}}]{jenkins18}
{Jenkins}, J.~M., {Long}, D.~M., {van Driel-Gesztelyi}, L., \& {Carlyle}, J.
  2018, \solphys, 293, 7

\bibitem[{{Jing} {et~al.}(2010){Jing}, {Yuan}, {Wiegelmann}, {Xu}, {Liu}, \&
  {Wang}}]{jing10}
{Jing}, J., {Yuan}, Y., {Wiegelmann}, T., {et~al.} 2010, \apjl, 719, L56

\bibitem[{{Jing} {et~al.}(2004){Jing}, {Yurchyshyn}, {Yang}, {Xu}, \&
  {Wang}}]{jing04}
{Jing}, J., {Yurchyshyn}, V.~B., {Yang}, G., {Xu}, Y., \& {Wang}, H. 2004,
  \apj, 614, 1054

\bibitem[{{Jur{\v{c}}{\'a}k} {et~al.}(2019){Jur{\v{c}}{\'a}k}, {Collados},
  {Leenaarts}, {van Noort}, \& {Schlichenmaier}}]{EST}
{Jur{\v{c}}{\'a}k}, J., {Collados}, M., {Leenaarts}, J., {van Noort}, M., \&
  {Schlichenmaier}, R. 2019, Advances in Space Research, 63, 1389

\bibitem[{{Kentischer} {et~al.}(2008){Kentischer}, {Bethge}, {Elmore},
  {Friedlein}, {Halbgewachs}, {Kn{\"o}lker}, {Peter}, {Schmidt}, {Sigwarth}, \&
  {Streander}}]{kentischer08}
{Kentischer}, T.~J., {Bethge}, C., {Elmore}, D.~F., {et~al.} 2008, in
  \procspie, Vol. 7014, Ground-based and Airborne Instrumentation for Astronomy
  II. Edited by McLean, Ian S.; Casali, Mark M. Proceedings of the SPIE, Volume
  7014, article id. 701413, 11 pp. (2008)., 701413

\bibitem[{{Kleint} {et~al.}(2015){Kleint}, {Battaglia}, {Reardon}, {Sainz
  Dalda}, {Young}, \& {Krucker}}]{kleint15}
{Kleint}, L., {Battaglia}, M., {Reardon}, K., {et~al.} 2015, \apj, 806, 9

\bibitem[{{Kliem} \& {T{\"o}r{\"o}k}(2006)}]{kliem06}
{Kliem}, B. \& {T{\"o}r{\"o}k}, T. 2006, \prl, 96, 255002

\bibitem[{{Kuckein} {et~al.}(2009){Kuckein}, {Centeno}, {Mart{\'{\i}}nez
  Pillet}, {Casini}, {Manso Sainz}, \& {Shimizu}}]{kuckein09}
{Kuckein}, C., {Centeno}, R., {Mart{\'{\i}}nez Pillet}, V., {et~al.} 2009,
  \aap, 501, 1113

\bibitem[{{Kuckein} {et~al.}(2017){Kuckein}, {Denker}, {Verma}, {Balthasar},
  {Gonz{\'a}lez Manrique}, {Louis}, \& {Diercke}}]{stools}
{Kuckein}, C., {Denker}, C., {Verma}, M., {et~al.} 2017, in IAU Symposium, Vol.
  327, Fine Structure and Dynamics of the Solar Atmosphere, ed. S.~{Vargas
  Dom{\'{\i}}nguez}, A.~G. {Kosovichev}, P.~{Antolin}, \& L.~{Harra}, 20--24

\bibitem[{{Kuckein} {et~al.}(2012){Kuckein}, {Mart{\'{\i}}nez Pillet}, \&
  {Centeno}}]{kuckein12a}
{Kuckein}, C., {Mart{\'{\i}}nez Pillet}, V., \& {Centeno}, R. 2012, \aap, 539,
  A131

\bibitem[{{Lagg} {et~al.}(2004){Lagg}, {Woch}, {Krupp}, \& {Solanki}}]{lagg04}
{Lagg}, A., {Woch}, J., {Krupp}, N., \& {Solanki}, S.~K. 2004, \aap, 414, 1109

\bibitem[{{Lemen} {et~al.}(2012){Lemen}, {Title}, {Akin}, {Boerner}, {Chou},
  {Drake}, {Duncan}, {Edwards}, {Friedlaender}, {Heyman}, {Hurlburt}, {Katz},
  {Kushner}, {Levay}, {Lindgren}, {Mathur}, {McFeaters}, {Mitchell}, {Rehse},
  {Schrijver}, {Springer}, {Stern}, {Tarbell}, {Wuelser}, {Wolfson}, {Yanari},
  {Bookbinder}, {Cheimets}, {Caldwell}, {Deluca}, {Gates}, {Golub}, {Park},
  {Podgorski}, {Bush}, {Scherrer}, {Gummin}, {Smith}, {Auker}, {Jerram},
  {Pool}, {Soufli}, {Windt}, {Beardsley}, {Clapp}, {Lang}, \& {Waltham}}]{aia}
{Lemen}, J.~R., {Title}, A.~M., {Akin}, D.~J., {et~al.} 2012, \solphys, 275, 17

\bibitem[{{Li} {et~al.}(2017){Li}, {Su}, {Zhou}, {van Ballegooijen}, {Sun}, \&
  {Ji}}]{li17}
{Li}, S., {Su}, Y., {Zhou}, T., {et~al.} 2017, \apj, 844, 70

\bibitem[{{Low}(1996)}]{low96}
{Low}, B.~C. 1996, \solphys, 167, 217

\bibitem[{{Mackay} {et~al.}(2010){Mackay}, {Karpen}, {Ballester}, {Schmieder},
  \& {Aulanier}}]{mackay10}
{Mackay}, D.~H., {Karpen}, J.~T., {Ballester}, J.~L., {Schmieder}, B., \&
  {Aulanier}, G. 2010, \ssr, 151, 333

\bibitem[{{Merenda} {et~al.}(2006){Merenda}, {Trujillo Bueno}, {Landi
  Degl'Innocenti}, \& {Collados}}]{merenda06}
{Merenda}, L., {Trujillo Bueno}, J., {Landi Degl'Innocenti}, E., \& {Collados},
  M. 2006, \apj, 642, 554

\bibitem[{{Moon} {et~al.}(2002){Moon}, {Choe}, {Wang}, {Park}, {Gopalswamy},
  {Yang}, \& {Yashiro}}]{moon02}
{Moon}, Y.~J., {Choe}, G.~S., {Wang}, H., {et~al.} 2002, \apj, 581, 694

\bibitem[{{Moore} {et~al.}(2001){Moore}, {Sterling}, {Hudson}, \&
  {Lemen}}]{moore01}
{Moore}, R.~L., {Sterling}, A.~C., {Hudson}, H.~S., \& {Lemen}, J.~R. 2001,
  \apj, 552, 833

\bibitem[{{Muglach} {et~al.}(1997){Muglach}, {Schmidt}, \&
  {Knoelker}}]{muglach97}
{Muglach}, K., {Schmidt}, W., \& {Knoelker}, M. 1997, \solphys, 172, 103

\bibitem[{{Panos} {et~al.}(2018){Panos}, {Kleint}, {Huwyler}, {Krucker},
  {Melchior}, {Ullmann}, \& {Voloshynovskiy}}]{panos18}
{Panos}, B., {Kleint}, L., {Huwyler}, C., {et~al.} 2018, \apj, 861, 62

\bibitem[{{Parenti}(2014)}]{parenti14}
{Parenti}, S. 2014, Living Reviews in Solar Physics, 11, 1

\bibitem[{{Penn}(2000)}]{penn00}
{Penn}, M.~J. 2000, \solphys, 197, 313

\bibitem[{{Penn} \& {Kuhn}(1995)}]{penn95}
{Penn}, M.~J. \& {Kuhn}, J.~R. 1995, \apjl, 441, L51

\bibitem[{{Pesnell} {et~al.}(2012){Pesnell}, {Thompson}, \& {Chamberlin}}]{sdo}
{Pesnell}, W.~D., {Thompson}, B.~J., \& {Chamberlin}, P.~C. 2012, Sol.\ Phys.,
  275, 3

\bibitem[{{Pietarila} {et~al.}(2007){Pietarila}, {Socas-Navarro}, \&
  {Bogdan}}]{pietarila07}
{Pietarila}, A., {Socas-Navarro}, H., \& {Bogdan}, T. 2007, \apj, 663, 1386

\bibitem[{{P{\"o}tzi} {et~al.}(2013){P{\"o}tzi}, {Hirtenfellner-Polanec}, \&
  {Temmer}}]{potzi13}
{P{\"o}tzi}, W., {Hirtenfellner-Polanec}, W., \& {Temmer}, M. 2013, Central
  European Astrophysical Bulletin, 37, 655

\bibitem[{{Robustini} {et~al.}(2019){Robustini}, {Esteban Pozuelo},
  {Leenaarts}, \& {de la Cruz Rodr{\'\i}guez}}]{robustini19}
{Robustini}, C., {Esteban Pozuelo}, S., {Leenaarts}, J., \& {de la Cruz
  Rodr{\'\i}guez}, J. 2019, \aap, 621, A1

\bibitem[{{Sainz Dalda} {et~al.}(2019){Sainz Dalda}, {de la Cruz
  Rodr{\'\i}guez}, {De Pontieu}, \& {Go{\v{s}}i{\'c}}}]{sainz_dalda2019}
{Sainz Dalda}, A., {de la Cruz Rodr{\'\i}guez}, J., {De Pontieu}, B., \&
  {Go{\v{s}}i{\'c}}, M. 2019, \apjl, 875, L18

\bibitem[{{Sasso} {et~al.}(2011){Sasso}, {Lagg}, \& {Solanki}}]{sasso11}
{Sasso}, C., {Lagg}, A., \& {Solanki}, S.~K. 2011, \aap, 526, A42

\bibitem[{{Sasso} {et~al.}(2014){Sasso}, {Lagg}, \& {Solanki}}]{sasso14}
{Sasso}, C., {Lagg}, A., \& {Solanki}, S.~K. 2014, \aap, 561, A98

\bibitem[{{Schad} {et~al.}(2016){Schad}, {Penn}, {Lin}, \& {Judge}}]{schad16}
{Schad}, T.~A., {Penn}, M.~J., {Lin}, H., \& {Judge}, P.~G. 2016, \apj, 833, 5

\bibitem[{{Scherrer} {et~al.}(2012){Scherrer}, {Schou}, {Bush}, {Kosovichev},
  {Bogart}, {Hoeksema}, {Liu}, {Duvall}, {Zhao}, {Title}, {Schrijver},
  {Tarbell}, \& {Tomczyk}}]{hmi}
{Scherrer}, P.~H., {Schou}, J., {Bush}, R.~I., {et~al.} 2012, Sol.\ Phys., 275,
  207

\bibitem[{{Schmidt} {et~al.}(2012){Schmidt}, {von der L{\"u}he}, {Volkmer},
  {Denker}, {Solanki}, {Balthasar}, {Bello Gonzalez}, {Berkefeld}, {Collados},
  {Fischer}, {Halbgewachs}, {Heidecke}, {Hofmann}, {Kneer}, {Lagg}, {Nicklas},
  {Popow}, {Puschmann}, {Schmidt}, {Sigwarth}, {Sobotka}, {Soltau}, {Staude},
  {Strassmeier}, \& {Waldmann }}]{schmidt12}
{Schmidt}, W., {von der L{\"u}he}, O., {Volkmer}, R., {et~al.} 2012, AN, 333,
  796

\bibitem[{{Sterling} \& {Moore}(2004)}]{sterling04}
{Sterling}, A.~C. \& {Moore}, R.~L. 2004, \apj, 613, 1221

\bibitem[{{Sterling} \& {Moore}(2005)}]{sterling05}
{Sterling}, A.~C. \& {Moore}, R.~L. 2005, \apj, 630, 1148

\bibitem[{{Tandberg-Hanssen}(1995)}]{tandberg95}
{Tandberg-Hanssen}, E. 1995, {The nature of solar prominences}, Vol. 199

\bibitem[{{T{\"o}r{\"o}k} \& {Kliem}(2005)}]{torok05}
{T{\"o}r{\"o}k}, T. \& {Kliem}, B. 2005, \apjl, 630, L97

\bibitem[{{Tritschler} {et~al.}(2016){Tritschler}, {Rimmele}, {Berukoff},
  {Casini}, {Kuhn}, {Lin}, {Rast}, {McMullin}, {Schmidt}, {W{\"o}ger}, \&
  {DKIST Team}}]{DKIST}
{Tritschler}, A., {Rimmele}, T.~R., {Berukoff}, S., {et~al.} 2016,
  Astronomische Nachrichten, 337, 1064

\bibitem[{{Trujillo Bueno} {et~al.}(2002){Trujillo Bueno}, {Landi
  Degl'Innocenti}, {Collados}, {Merenda}, \& {Manso Sainz}}]{trujillo02}
{Trujillo Bueno}, J., {Landi Degl'Innocenti}, E., {Collados}, M., {Merenda},
  L., \& {Manso Sainz}, R. 2002, \nat, 415, 403

\bibitem[{{van Ballegooijen} \& {Martens}(1989)}]{vanballe89}
{van Ballegooijen}, A.~A. \& {Martens}, P.~C.~H. 1989, \apj, 343, 971

\bibitem[{{Vial} \& {Engvold}(2015)}]{vial15}
{Vial}, J.-C. \& {Engvold}, O. 2015, {Solar Prominences}, Vol. 415

\bibitem[{{Viticchi{\'e}} \& {S{\'a}nchez Almeida}(2011)}]{viticchie11}
{Viticchi{\'e}}, B. \& {S{\'a}nchez Almeida}, J. 2011, \aap, 530, A14

\bibitem[{{von der L{\"u}he}(1998)}]{vonderluehe98}
{von der L{\"u}he}, O. 1998, {New Astron. Rev.}, 42, 493

\bibitem[{{Wang} {et~al.}(2020){Wang}, {Jenkins}, {Martinez Pillet}, {Beck},
  {Long}, {Prasad Choudhary}, {Muglach}, \& {McAteer}}]{wang20}
{Wang}, S., {Jenkins}, J.~M., {Martinez Pillet}, V., {et~al.} 2020, \apj, 892,
  75

\bibitem[{{Xu} {et~al.}(2012){Xu}, {Lagg}, {Solanki}, \& {Liu}}]{xu12}
{Xu}, Z., {Lagg}, A., {Solanki}, S., \& {Liu}, Y. 2012, \apj, 749, 138

\bibitem[{{Xue} {et~al.}(2016){Xue}, {Yan}, {Cheng}, {Yang}, {Su}, {Kliem},
  {Zhang}, {Liu}, {Bi}, {Xiang}, {Yang}, \& {Zhao}}]{xue16}
{Xue}, Z., {Yan}, X., {Cheng}, X., {et~al.} 2016, Nature Communications, 7,
  11837

\bibitem[{{Yan} {et~al.}(2020){Yan}, {Xue}, {Cheng}, {Zhang}, {Wang}, {Kong},
  {Yang}, {Chen}, \& {Feng}}]{yan20}
{Yan}, X., {Xue}, Z., {Cheng}, X., {et~al.} 2020, \apj, 889, 106

\bibitem[{{Yardley} {et~al.}(2018){Yardley}, {Green}, {van Driel-Gesztelyi},
  {Williams}, \& {Mackay}}]{yardley18}
{Yardley}, S.~L., {Green}, L.~M., {van Driel-Gesztelyi}, L., {Williams}, D.~R.,
  \& {Mackay}, D.~H. 2018, \apj, 866, 8

\bibitem[{{Yelles Chaouche} {et~al.}(2012){Yelles Chaouche}, {Kuckein},
  {Mart{\'{\i}}nez Pillet}, \& {Moreno-Insertis}}]{yelles12}
{Yelles Chaouche}, L., {Kuckein}, C., {Mart{\'{\i}}nez Pillet}, V., \&
  {Moreno-Insertis}, F. 2012, \apj, 748, 23

\bibitem[{{Zhang} {et~al.}(2008){Zhang}, {Zhang}, \& {Zhang}}]{zhang08}
{Zhang}, Y., {Zhang}, M., \& {Zhang}, H. 2008, \solphys, 250, 75

\end{thebibliography}

%--------------------------------------------------------------------------
% Appendix
%--------------------------------------------------------------------------

%\appendix
\begin{appendix}

\section{{\it k}-means clustering}\label{App:A}

\begin{figure}[!h]
\centering

\includegraphics[width=1.0\textwidth]{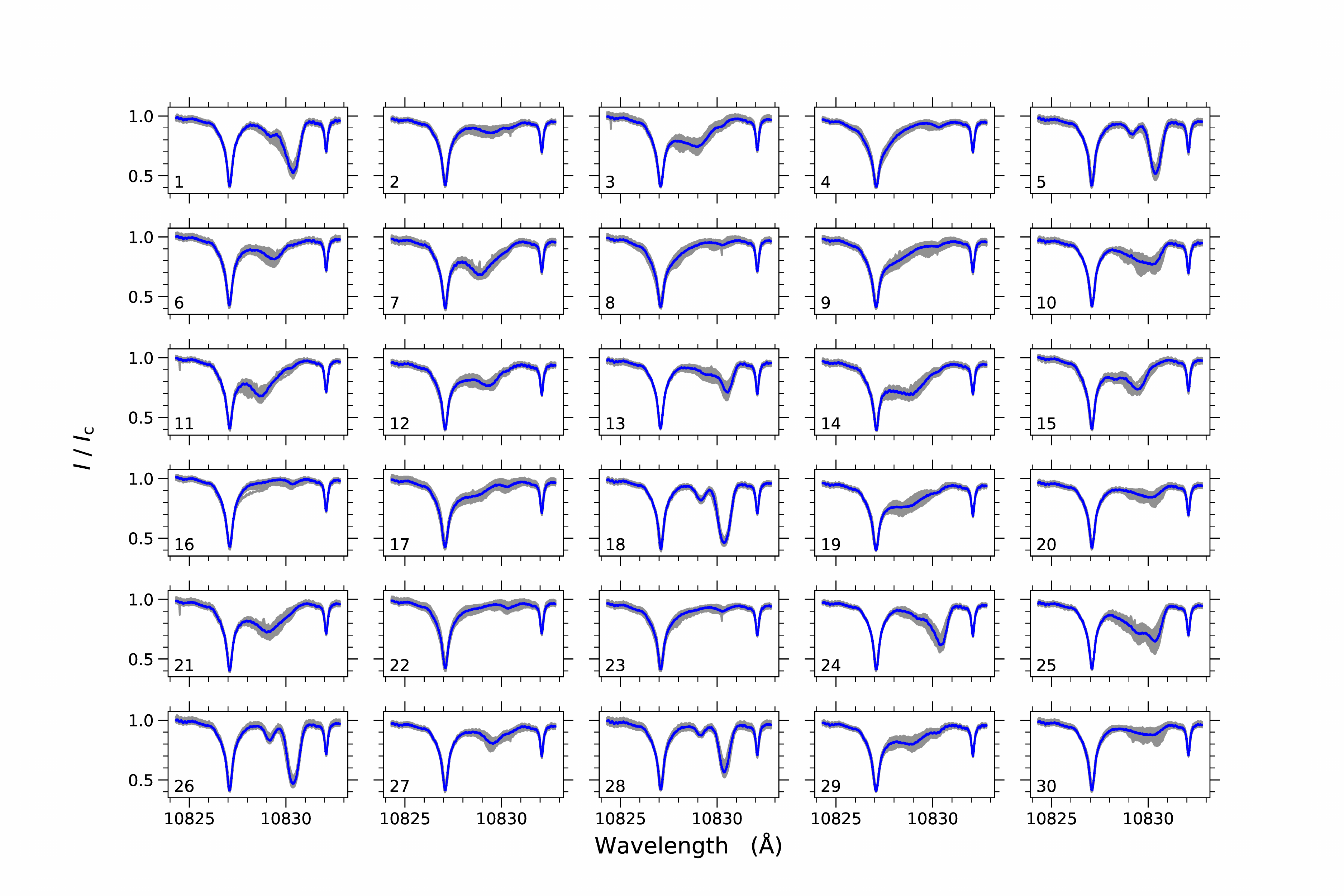}
\caption{Classification of the Stokes $I$ profiles within map A using $k$-means. A total amount
of 30 cluster represent well the variety of intensity profiles across the map. 
The blue profile is the average of 
all the individual profiles, which appear in gray color, within one cluster. Each cluster is
identified by a number, which appears in the lower left corner of each panel.} 
\label{Fig:cluster30}

\end{figure}

An overview of all the 30 groups obtained from the $k$-means
clustering is shown in Fig. \ref{Fig:cluster30}.

%---- Figure X -----------------------------------------------------------------
%-------------------------------------------------------------------------------

\end{appendix}

\end{document}